\documentclass[review]{elsarticle}

\usepackage{lineno,hyperref}
\modulolinenumbers[5]

\usepackage{amsmath,amssymb,amsfonts}
\usepackage{algorithmic}
\usepackage{graphicx}
\usepackage{textcomp}
\usepackage{xcolor}
\usepackage{hyperref}
\usepackage{multirow}
\usepackage{float}
\usepackage{pgf-pie}
\usepackage{longtable}    
\usepackage{caption}
\usepackage{subcaption}

\usepackage{xcolor}
\newcommand{\red}[1]{\textcolor{black}{#1}}

\journal{Elsevier - Internet of Things}

\begin{document}

\begin{frontmatter}
\title{MIMA 2.0 - \textcolor{black}{Compact and Portable} \\
Multifunctional IoT integrated Menstrual Aid}

\author{Jyothish Kumar J\corref{mycorrespondingauthor}}
\cortext[mycorrespondingauthor]{Corresponding author}
\ead{jyothishk.j@niser.ac.in}
\author{Subhankar Mishra}

\address{National Institute of Science Education and Research, Bhubaneswar}

\author{Shreya Shivangi}
\author{Amish Bibhu}
\author{Sulagna Saha}
\address{National Institute of Fashion Technology, Bhubaneswar}




\begin{abstract}

The shredding \red{intrauterine} lining or the endometrium is known as Menstruation. It occurs every month and causes several issues like Menstrual Cramps and ache\red{s} in \red{the} abdominal region, stains, menstrual malodor, rashes in intimate areas\red{,} and many more. In our research, almost \textcolor{black}{all} of the products available in the market do not cater to these problems single-handedly. There are few remedies available to cater to the cramps, among which heat therapy is the most commonly used. Our methodology, involved surveys regarding problems and the solutions to these problems that \textcolor{black}{are deemed} optimal. This inclusive approach helped us infer about the gaps in available menstrual aids which has become our guide towards developing MIMA (Multifunctional IoT Integrated Menstrual Aid). In this paper, we have featured an IOT incorporated multifunctional smart intimate wear that aims to provide for the multiple necessities of women during menstruation like leakproof, antibacterial, anti-odor, rash-free experience along with an integrated Bluetooth\red{-}controlled intimate heat-pad for rel\red{ie}ving abdominal cramps. The entire process of product development has been done in phases according to feedback from target users in each stage. 
This paper is an extension to our paper \cite{mima1} which serves as the proof of concept for our approach. \textcolor{black}{The} development has led us towards MIMA 2.0 featuring a completely \textcolor{black}{con\red{c}ealed and} integrated design that includes a safe Bluetooth-controlled heating system for the intimate area. The product has received incredibly positive feedback from survey participants.

\end{abstract}

\begin{keyword}
\texttt \ {Menstruation}\sep \ {Period Pants}\sep \ {Smart wear}\sep {Dysmenorrhea} \sep\ {IoT}\sep\ {intimate wear}\sep
\end{keyword}

\end{frontmatter}

\linenumbers

\section{Introduction}

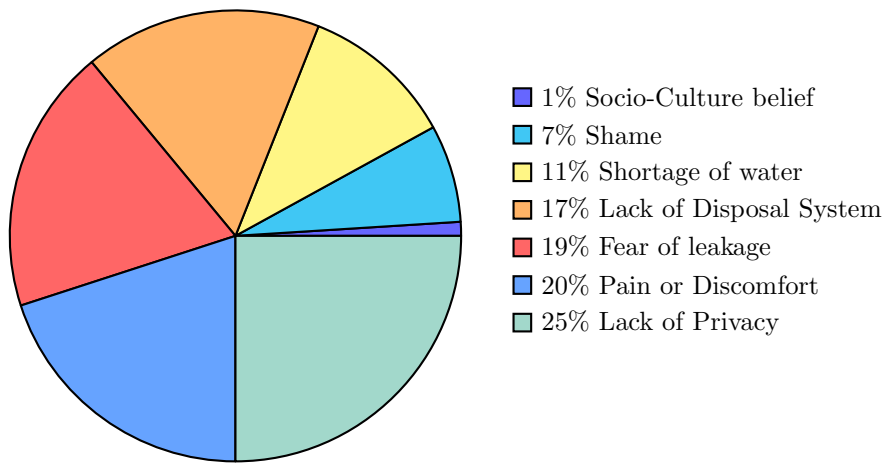
\begin{figure}[H]
    \centering
\begin{tikzpicture}
\pie[text=legend, hide number]{1/1\% Socio-Culture belief, 7/7\% Shame, 11/11\% Shortage of water, 17/17\% Lack of Disposal System, 19/19\% Fear of leakage, 20/20\% Pain or Discomfort, 25/25\% Lack of Privacy}
\end{tikzpicture}
    \caption{Reason of Absenteeism.}
    \label{fig:Absenteeism}
\end{figure}

Although most women have regular menstrual cycles, there is no denying the fact that many women face multiple irregularities and problems during menstruation. Many adolescent girls  are constantly afraid of leakage spots throughout their cycles and suffer  from multiple cramps. These cramps are symptoms of irregular menstruation known as dysmenorrhea. It can be divided into primary dysmenorrhea and secondary dysmenorrhea. Primary dysmenorrhea is the most common recurrent menstrual pain in women. Cramps and pain usually begin 2-3 days before bleeding and may continue for the first 3-4 days of the cycle. Recurrent cramps and fatigue can also be felt in the abdomen, lower back, hips, and thighs. Secondary dysmenorrhea can be the result of multiple problems in the reproductive system, including endometriosis, pelvic inflammatory disease, adenomyosis, fibroids, etc \cite{oladosu2018abdominal}. Constant use of the pads also causes skin rashes caused by the constant friction between the delicate inner thighs and the extra microplastic on the pad wings. Some of the major reasons for being absent from school are discussed in Figure \ref{fig:Absenteeism}.


This paper covers the design and staged development \cite{mima1} of a new class of menstrual aid for women. The paper presents our research into the area of menstrual problems and available aids and remedial products for women. Through this paper, we also present our initiative to fulfill the need for the development of a new age IoT-powered period pants \red{that go} beyond conventional forms of menstrual support by incorporating a safe and intimate heating method for menstrual cramps experienced by women among other side effects of menstruation. \\


MIMA 2.0 \textcolor{black}{is evolved from our proof of concept - MIMA 1.0 based on the user \red{feedback} obtained from fifty female volunteers. Their inputs on the design and implementation of the concept carved our way towards MIMA 2.0 which is designed as an affordable solution for multiple issues faced by women during menstruation. MIMA 2.0 }  makes changes to the construction of the module, enabling key features such as improvising a leak-proof gusset, in addition to a mini pocket to house the extra part of the sanitary napkin wings to prevent rashes. An IoT intervention is performed to integrate a very slim and light thermal pad placed in the internal pocket of the abdomen.

The rest of the paper is \red{organized} as follows. Section \ref{sec:problem} describes the problem statement, it covers the fundamental terminologies associated with \red{the} paper including - Menstruation, Early onset of Menarche, Menstrual discomfort, Menstrual cramps\red{,} and remedies. Section \ref{sec:relatedworks} summarizes various studies \red{on} the problems associated with menstruation and the commercial initiatives aimed at addressing these issues. The next section \ref{sec:methodology} covers the details of the survey and analysis that formed the base for the development of MIMA. The section covers the process of \red{multi-stage} development of MIMA underpants garment design and the IoT-controlled heating system of MIMA. This is followed by the result and analysis of our work in Section \ref{sec:result} and \red{the} conclusion of the paper in \ref{sec:conclusion}. \textcolor{black}{Please note that this paper refers to our proof of concept as MIMA 1.0. The MIMA 1.0 and \red{its} user feedback \red{were} presented in our previous publication. The current paper presents the evolution of the current version of MIMA (viz. MIMA 2.0) through constant development while working with volunteering target beneficiaries. }

\begin{figure}[h!]
     \centering
     \includegraphics[width=\linewidth]{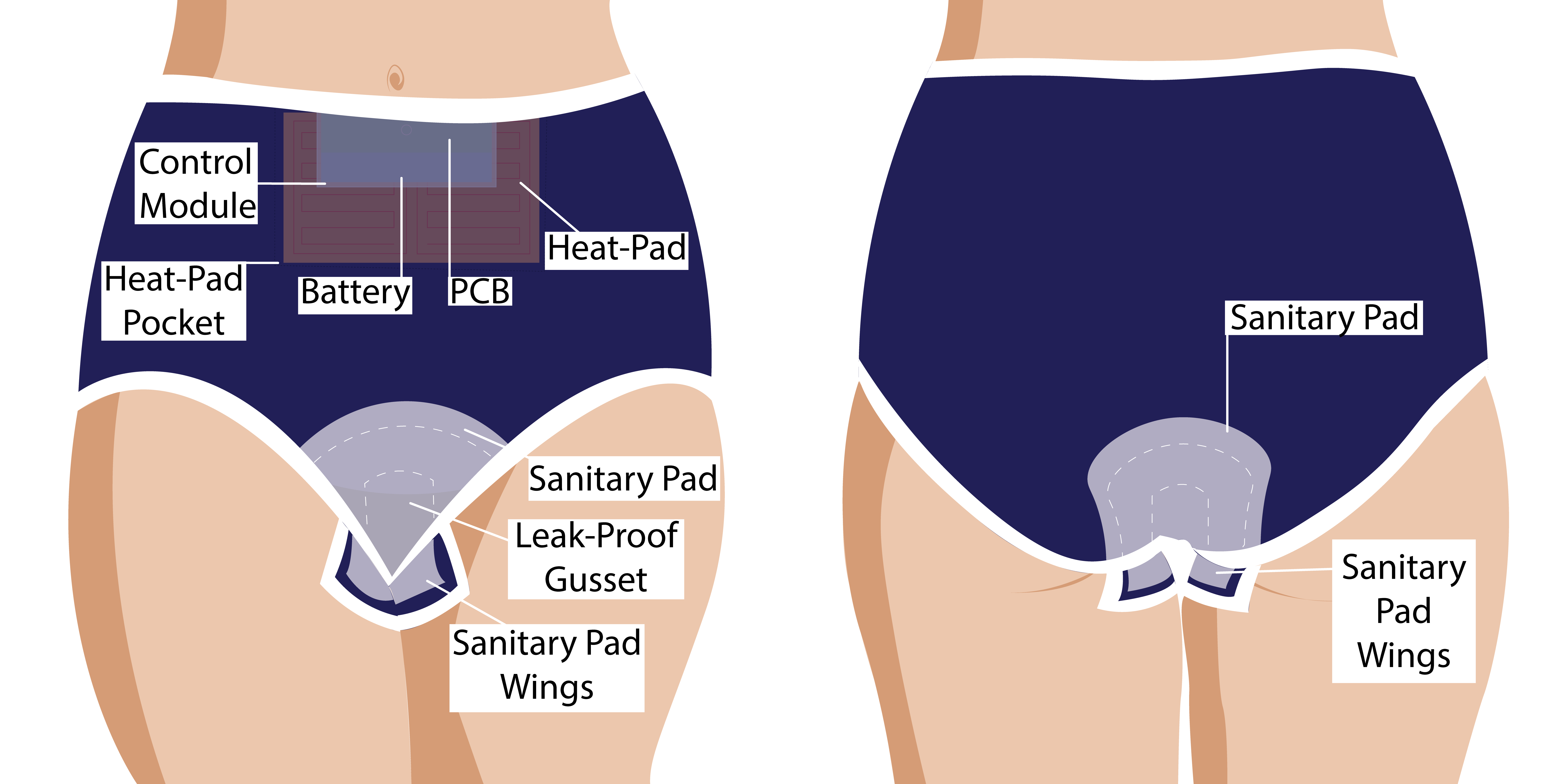}
    \caption{Illustration of design and placement of MIMA 2.0 and over body-form illustration.}
    \label{fig:MIMA2Placement}
    
\end{figure}

\begin{table}[]
    \centering
   \caption{Feature comparison of MIMA 1.0 and MIMA 2.0.}
   \label{tab:featurecomp}
\small
\begin{longtable}{|p{0.2\textwidth}|p{0.35\textwidth}|p{0.35\textwidth}|}
    \hline
    \textbf{Feature}& \textbf{MIMA 1.0 }&\textbf{MIMA 2.0}\\
     \hline
     \endhead
\textcolor{black}{Status} & \textcolor{black}{Developed as a proof of concept} & \textcolor{black}{The current state of the product.}\\

\hline
Weight  & 210 Gm  & 65 Gm  \\

\hline
Dimensions  & 10x6x3.8 cm  & 10x5x0.8 cm \\

\hline
Battery Capacity  &  2200 mAh & 2200 mAh  \\

\hline
Operating Voltage & 11.2 Volt to 12.6 Volt & 3.7 Volt to 4.2 Volt  \\

\hline
Heating element & 32 AWG Nichrome (52 cm each.) & 36 AWG Nichrome (16 cm each.)\\

\hline
Max Operating Current & 2.8 Amps & 5 Amps  \\

\hline
\red{Continuous} runtime at Max Power  &  1.5 Hours & 0.5 Hours  \\

\hline
\red{Comparative} Feedback Analysis &  \begin{itemize} \item Concept   is great \item Problem statement is valid and MIMA is effective \item Significant help   in menstrual cramps \item It'll be easier if the connection to the heating pad is   made removable \item \red{The} external nature of the controller module is a concern\end{itemize}  & \begin{itemize} \item It's   slicker now \item The removable connection is much better \item User experience   otherwise is \red{the} same \item \red{The effectiveness} of MIMA is \red{reemphasized}.\end{itemize}  \\

\hline

\end{longtable}

\end{table}

\section{Problem Statement} 
\label{sec:problem}

\subsection{Menstruation}

\begin{figure}[h]
     \centering
     \includegraphics[width=\linewidth]{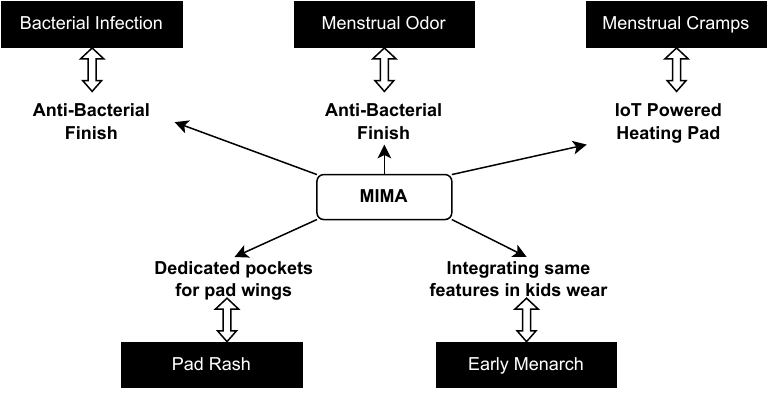}
    \caption{Problems related to menstruation and solutions incorporated in MIMA.}
    \label{tab:solutions}
\end{figure}

Menstruation is a very critical and complex mechanism in a woman’s body that depends not just on reproductive organs but multiple hormones like \red{estrogen} \& progesterone, food habits, emotional and physical well-being, heredity, and a lot more. It is the removal of the endometrium lining of the uterus through the vagina. The menstrual fluid consists of blood, cells from the uterus endometrium lining, and mucus. The average length of menstruation is usually between three to seven days but it can be of a one-day length to as long as eight days normally in a female. Pain-free menstruation and a regular menstrual cycle would require a perfect synchronization of all these factors \cite{thiyagarajan2020physiology}. \textcolor{black}{Figure} \ref{tab:solutions} shows the solutions we are working on in MIMA to resolve the issues faced during menstruation. 

\subsection{Early onset of Menarche and \red{menstrual} discomfort}
\red{The} typical age for menarche or the first menstruation of a wom\red{a}n’s life is 12-14 years. Menarche marks the development of sexual characteristics in girls. There has been a rise in early menarche, at the age as low as 8–10 years lately. There are many reasons for this including consumption of genetically engineered fruits and vegetables, presence of pesticides in food, increased poultry diet\red{,} etc. \red{A} sedentary lifestyle and obesity in children can also cause physical and mental stress which in turn may trigger early menarche. \cite{dharmarha2018study}. \\

This onset of menstruation at this early age can make it difficult for the person to cope with menstrual cramps, mood swings\red{,} and drastic physical changes \cite{grandi2012prevalence} which may pose difficulty in handling school and studies. Likewise\red{,} the same issues of cramps, body aches, headaches, fear of spotting and leakage\red{,} etc. affect adult women also\red{,} often impacting their work. Another prominent issue is that of menstrual malodor which is an unusual fishy odor caused by the presence of bacterium \emph{Gardenerella vaginalis} in vaginal discharge.

\subsection{Menstrual Cramps and Remedies}
Menstrual cramps are caused due to extra visceromotor reflexes in relation to \red{the} shedding of the endometrial lining \cite{ oladosu2018abdominal}. Painkillers, heat therapy\red{,} and acupressure\red{,} etc. are some of the remedies that women go to for comforting their constant pain. Hot water bags are \red{the} most accepted and efficient pain relief methods that are often recommended by medical practitioners. According to research, applying heat at 40-45 degree\red{s} Celsius at 1-2 cm depth in \red{the} pelvic and abdominal region stimulates increased blood circulation, thereby resolving fluid retention and the resultant swelling and pain caused due to nerve congestion. \cite{ higgins2002quantifying}. Despite the efficiency of heat therapy, it is not possible for women to benefit from \red{it} while in school or at work. \textcolor{black}{MIMA is the potential solution in these situations.}

\section{Related Works}
\label{sec:relatedworks}

\begin{table}[h]
\caption{Issues and existing solutions \cite{peberdy2019study} \cite{healthshots}.}\label{tab2}
\centering
\begin{tabular}{|p{0.2\linewidth}|p{0.7\linewidth}|}
\hline
\textbf{Issues}                                    & \textbf{Existing Solutions}                                     \\ \hline
\multirow{4}{*}{Cramp}                             & 1. Electric heating pads                                        \\ \cline{2-2} 
                                                   & 2. Hot Water Bag                                                \\ \cline{2-2} 
 & 3. Nua Heating Patches (An Eco-Friendly solution for cramp relief to be applied in the abdominal region) \\ \cline{2-2} 
                                                   & 4. Livia (Acupressure solution for cramp relief using impulses) \\ \hline
\multirow{4}{*}{Stain \& Leakage} & 1. Sanitary Pads                                                \\ \cline{2-2} 
                                                   & 2. Tampons                                                      \\ \cline{2-2} 
                                                   & 3. Menstrual Cup                                                \\ \cline{2-2} 
                                                   & 4. Period Pants                                                 \\ \hline
\multirow{3}{*}{Odor}                              & 1. Scented Sanitary Pads                                        \\ \cline{2-2} 
                                                   & 2. Anti-bacterial finish Intimate Wear                           \\ \cline{2-2} 
                                                   & 3. Essential Oil                                                \\ \hline
Rashes                                             & Rash cure with ointments                                        \\ \hline
Reminder                                           & Flow App and other Period tracker apps                          \\ \hline
\end{tabular}
\end{table}

 Period Pants have been launched by different brands to provide women \red{with} a pad-free period whereas, the majority of women of menstrual age use sanitary pads only\cite{smith2020national} \cite{choi2021use}. The \red{work} aims at achieving the high absorption capacity of multiple pads/ tampons secured with a leak-proof outer layer. From our survey, we found out that very few people are using period pants, and most of the research is not oriented toward the Indian market. The current research is being done for women/girls starting from the age of 12, considering the starting age for Menstrual cycles, but in the current scenario, girls have started to get their first menstrual cycle before the age of 12 also. \cite{ruble1982experience} There are multiple products and works to solve these problems individually, as shown in Table \ref{tab2} but there is a gray area identified in the development of a product with all-round features.
 
The period pants available currently, have three layers: 
\begin{enumerate}
    \item \red{S}uper-dry layer at the top.
    \item \red{H}igh absorbent layer in the middle capable of absorbing multiple numbers of sanitary pads/tampons.
    \item \red{T}he last layer facing the main garment is the leak-proof layer.
\end{enumerate}

Preferability and acceptance of period pants \red{are} due to \red{the} economic factor and eco-friendly reusability. A woman spends around \$10,000- \$15000 on menstrual aids during their lifetime. Switching to reusable period pants would result in measurable financial savings. Surveys indicate that though women are open to period pants as a viable option to menstrual sanitary aid, In practice very low number of women are actually adopting them.\cite{phan2020acceptance}. There is quite a lot of motivation and awareness about period pants, but preferability is a lot less. A qualitative study outlines the values the surveyees place on the importance of proper sanitary infrastructure and menstrual hygiene education while in a refugee camp, it was a challenge to manage menstruation. Access to menstrual aids and their disposal is a major struggle. So, the dual use of reusable period pants and normal underwear was found to be very popularly preferable \cite{vanleeuwen2018exploring}. 

While we discuss different options \red{for} menstrual aids and talk about menstrual hygiene on multiple forums, there are still many rural households where women are not fortunate enough to be able to afford a pack of sanitary pads during menstruation, there are people who still use cloth pieces and have no option but to compromise their menstrual hygiene. \cite{kaur2018menstrual}


\section{Methodology}
\label{sec:methodology}

\subsection{Initial Data Collection and Analysis}
MIMA is targeted at menstruation-related problems such as abdominal cramps, leakage, pad rash, malodor, etc., faced by women. \textcolor{black}{We conducted an anonymous online survey to gather data for our research. The survey was designed to achieve specific objectives, and we ensured that no personal identifying information such as names or email IDs \red{was} collected. The survey was conducted among 170 subjects from India, representing women from diverse age groups (17 to 58) and professions. To reach out to potential participants, we shared the survey form on various social media platforms including WhatsApp, Instagram, and LinkedIn. The careful curation of the questionnaire allowed us to collect valuable insights from a wide range of participants, contributing to the robustness and reliability of our research findings.} These surveys helped us understand the neglected shortcomings of women's undergarments concerning menstruation, as well as the real-life problems associated with menstruation, along with information about popular approaches to remedy their discomforts. A summary of the survey responses that guided our initial study is shown in the figure. \ref{fig:surveysummary}.\\

\textcolor{black}{ As depicted, \red{w}omen prioritize fabric quality over other factors while buying intimate\textcolor{black}{ wear \ref{fig: buyingpref}}. \red{The} comfort of the \red{innerwear} is the most important feature for their intimate wear preference\red{s} \ref{fig:featurepreference}. Over 70 percent of the females in our survey group attained puberty during ages 12 -14 \ref{fig:startage}, \red{h}owever around 58.8 \% \red{of} females in our survey group know someone who attained menarche before the age of 12 years \ref{fig:before12}. 57.1 \% of women in our survey group suffer from some sort of menstrual irregularity \ref{fig:menstrualirregularities}. \red{The} majority (95.3 \%) of women use sanitary pads for their menstrual hygiene requirements \ref{fig:menstrualaid}, and around 56.5 \% of the women were aware of period pants as an alternative for menstrual aid \ref{fig:ppaware}. Figure \ref{fig:typescramps} shows the statistics of women experiencing different types of menstrual cramps. Out of the affected individuals, 58.8 \% \red{of} women use heat therapy as a remedy for their cramps \ref{fig:heatyn}. Figure \ref{fig:underweartype} shows the preference \red{for} style of underwear among our survey group. The preference \red{for} Hipster style underwear guided the development of MIMA around this type of underwear. The proposal of a solution like MIMA was gladly accepted by 76.5\% of women. This survey and analysis carved the way for our development in the initial stage of our product development which \red{led} to the proof of concept (MIMA 1.0), which further developed towards MIMA 2.0.}

\begin{figure}[H]
    \raggedright
    \begin{subfigure}[h]{0.45\textwidth}
    \raggedright
        \begin{tikzpicture}[scale=0.35]
        \pie[text=legend, hide number]{4.7/4.7\% Finish, 6.5/6.5 \% Price, 4.7/4.7\% Design, 9.4/9.4\% Brand, 74.7/74.7\% Fabric }
        \end{tikzpicture}
    \caption{Buying preference for intimate wear.}
    \label{fig: buyingpref}
    \end{subfigure}
    \hfill
    \begin{subfigure}[h]{0.45\textwidth}
         \raggedright
        \begin{tikzpicture}[scale=0.35]
        \pie[text=legend, hide number]{2.4/2.4\% Sustainability, 4.7/4.7\% Longevity, 2.4/2.4\% High Absorption, 5.3/5.3\% Natural Fibres, 85.3/85.3\% Comfort }
        \end{tikzpicture}
     \caption{Primary concerns while choosing an underwear.}
     \label{fig:featurepreference}
     \end{subfigure}
     \hfill \\
       \dotfill\\

     \begin{subfigure}[h]{0.45\textwidth}
     \raggedright
        \begin{tikzpicture}[scale=0.35]
        \pie[text=legend, hide number]{15.9/15.9\% Before 12, 70/70\% 12-14, 12.9/12.9\% 15-17, 1.2/1.2\% After 17}
        \end{tikzpicture}

    \caption{Age of attaining menarche.}
    \label{fig:startage}
    \end{subfigure}
    \hfill
    \begin{subfigure}[h]{0.45\textwidth}
    \raggedright
        \begin{tikzpicture}[scale=0.35]
        \pie[text=legend, hide number]{41.2/ 41.2\% No, 58.8/ 58.8\% Yes}
        \end{tikzpicture}

    \caption{Presence of a known female attaining menarche before 12 years.}
    \label{fig:before12}
    \end{subfigure}
    \hfill \\
    
    \dotfill\\

    \begin{subfigure}[h]{\textwidth}
     \raggedright
        \begin{tikzpicture}[scale=0.35]
        \pie[text=legend, hide number]{12.9/12.9\% Heavy Menstrual Bleeding, 2.9/2.9\% Amenorrhea (No Menstrual Bleeding), 25.3/25.3\% Dysmenorrhea (Painful Menstruation), 15.9/15.9\% Polycystic Ovary Syndrome (PCOS or PCOD), 42.9/42.9\% None of the Above}
        \end{tikzpicture}

    \caption{Menstrual irregularities faced.}
    \label{fig:menstrualirregularities}
    \end{subfigure}
    \hfill \\
      
    \dotfill\\

    \begin{subfigure}[h]{0.45\textwidth}
    \raggedright
        \begin{tikzpicture}[scale=0.35]
        \pie[text=legend, hide number]{95.3/95.3\% Sanitary Pads, 0.6/0.6\% Tampons, 4.1/4.1\% Menstrual Cups, 0/0\% Period Pants }
        \end{tikzpicture}
     \caption{Popularity of various menstrual aids.}
     \label{fig:menstrualaid}
     \end{subfigure}
     \hfill
    \begin{subfigure}[h]{0.45\textwidth}
    \raggedright
        \begin{tikzpicture}[scale=0.35]
        \pie[text=legend, hide number]{43.5/43.5\% No, 56.5/56.5\% Yes }
        \end{tikzpicture}
    \caption{Awareness about Period Pants.}
    \label{fig:ppaware}
    \end{subfigure}
     \hfill \\
   
   \dotfill\\     

        \caption{Summary of the survey.}
\label{fig:surveysummary}
\end{figure}
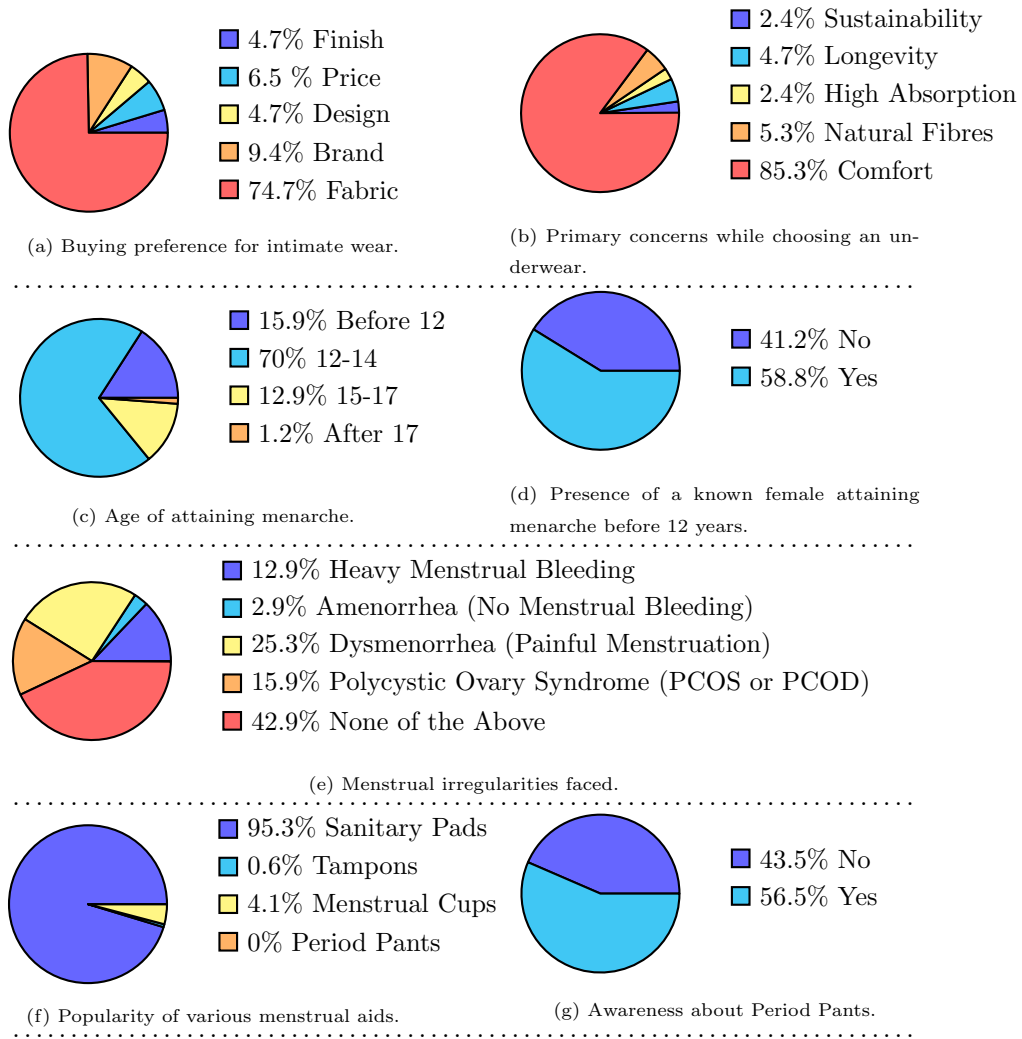

\begin{figure}
  \ContinuedFloat
\centering

    \begin{subfigure}[h]{0.45\textwidth}
    \raggedright
        \begin{tikzpicture}[scale=0.35]
        \pie[text=legend, hide number]{44.1/44.1\% Abdominal, 10.6/10.6\% Pelvic , 2.4/2.4\% Butt, 34.1/34.1\% All of these, 8.8/8.8\% None of these }
        \end{tikzpicture}
    \caption{Types of menstrual cramps experienced.}
    \label{fig:typescramps}
    \end{subfigure}
    \hfill
     \begin{subfigure}[h]{0.45\textwidth} 
    \raggedright
        \begin{tikzpicture}[scale=0.35]
        \pie[text=legend, hide number]{41.2/ 41.2\% No, 58.8/ 58.8\% Yes}
        \end{tikzpicture}
    \caption{Statistic of women using heat as cramp remedy.}
    \label{fig:heatyn}
    \end{subfigure}
    \hfill \\
   
\dotfill\\

    \begin{subfigure}[h]{0.45\textwidth}
    \raggedright
       \begin{tikzpicture}[scale=0.35]
        \pie[text=legend, hide number]{2.4/2.4\% Thong, 21.8/21.8\% Bikini, 27.6/27.6\% Hipser, 6.5/6.5\% High Briefs, 18.2/18.2\% Mid-Rise,15.9/15.9\% Cheekies, 7.6/7.6\% Boyshort }
        \end{tikzpicture}
    \caption{Popularity of various underwear types during menstruation.}
    \label{fig:underweartype}
    \end{subfigure}
    \hfill
    \begin{subfigure}[h]{0.45\textwidth}
    \raggedright
        \begin{tikzpicture}[scale=0.35]
        \pie[text=legend, hide number]{76.5/76.5\% Yes, 23.5/23.5\% No}
        \end{tikzpicture}
     \caption{Acceptability for concept of MIMA smart period pants.}
     \label{fig:acceptancemima}
     \end{subfigure}
     \hfill
\caption{Summary of the survey.}
\label{fig:surveysummary}
\end{figure}
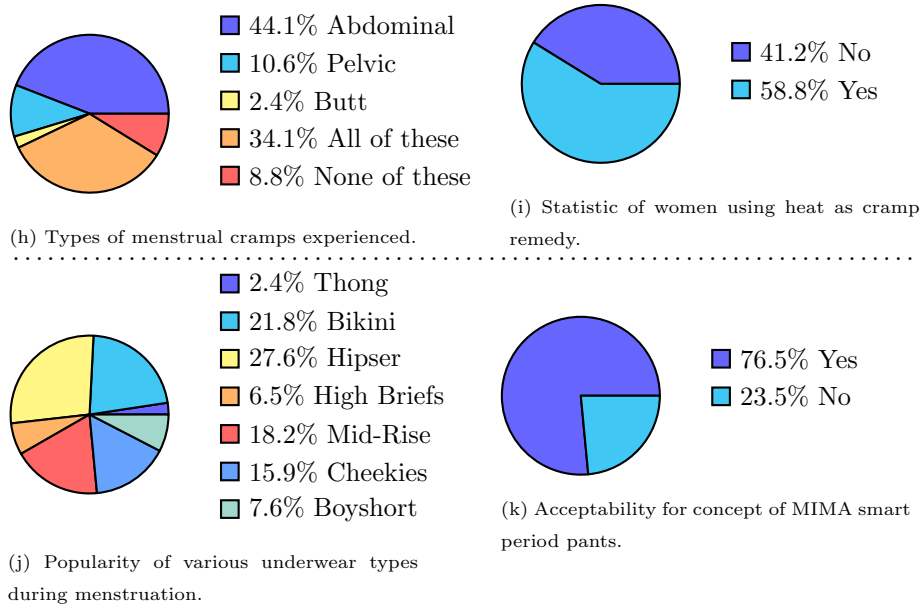

\subsection{Garment Design}

The first step in the development of MIMA was to design an undergarment that \red{could} address menstrual problems such as rashes, leakage, and malodor. The design of MIMA's clothing component was based on industry knowledge and the insights gained from our initial survey. The garment was manufactured using a 161 GSM blend of 93\% cotton and 7\% elastane, which is known to be optimal for intimate wear due to its natural fibers and soft feel.

The design of the garment is inspired by hipster-style women's underwear, with the waist modified to be high-waisted for better coverage. This modification allows for the accommodation of a removable IoT-powered heat pad that provides efficient heat distribution across the abdominal area. The waistband features a pocket designed to fully accommodate the slim and sleek IoT-powered heating pad and control module of MIMA.

\begin{figure}[H]
\centering
     \begin{subfigure}{0.45\textwidth}
         \centering
         \includegraphics[width=\textwidth]{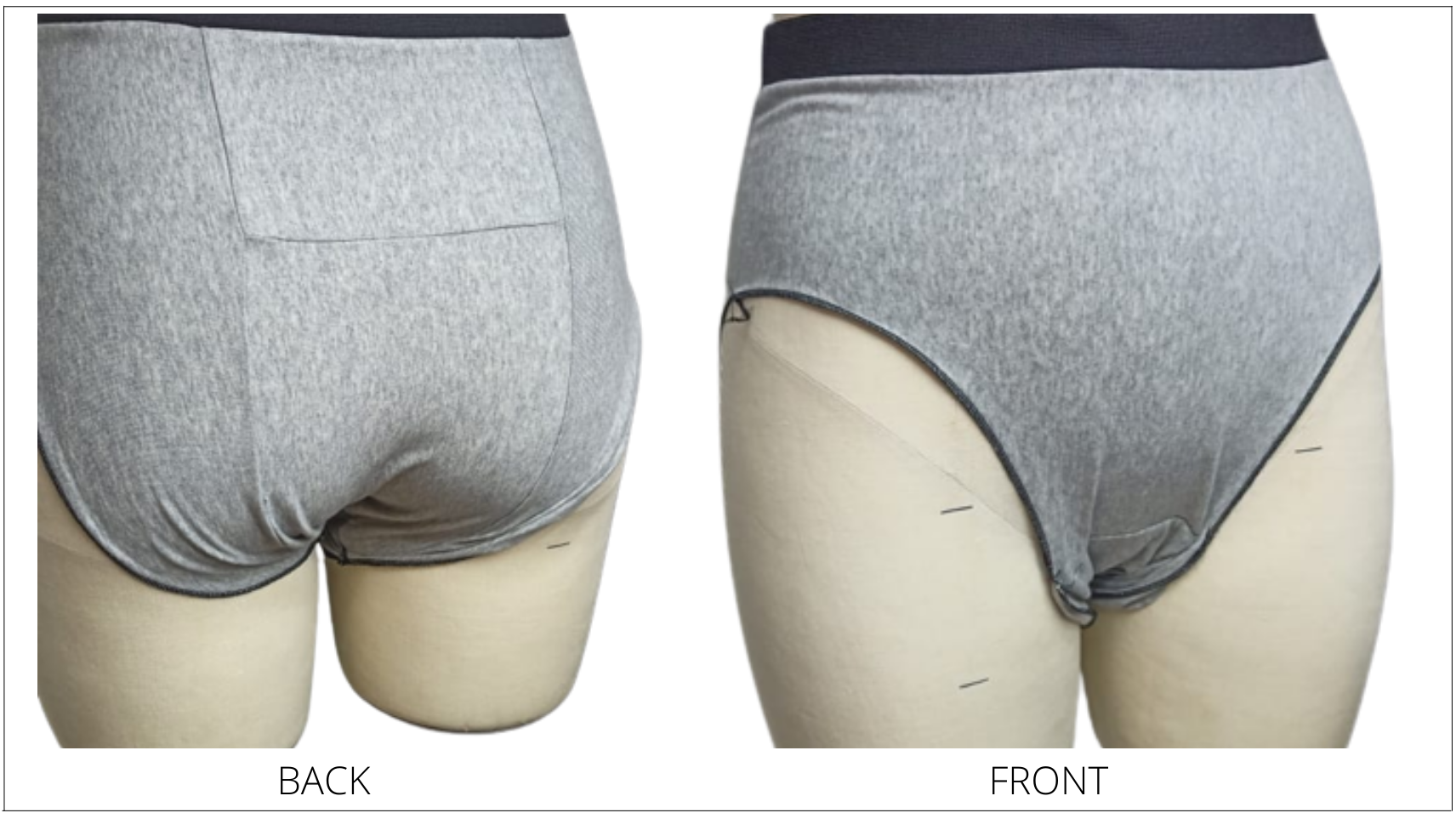}
         \caption{Proto Sample.}
         \label{fig: Proto Sample}
     \end{subfigure}
     \hfill
     \begin{subfigure}{0.45\textwidth}
         \centering
         \includegraphics[width=\textwidth]{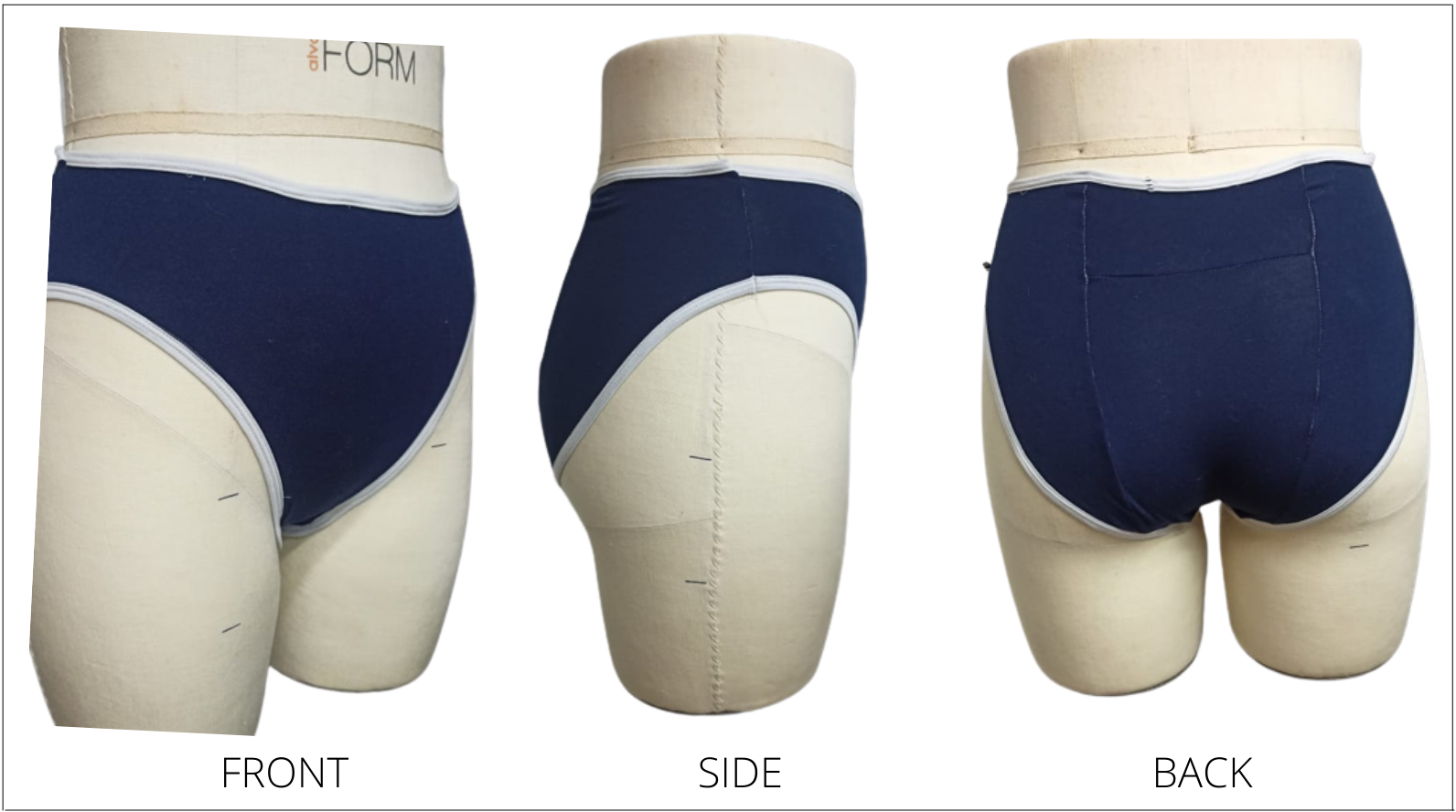}
         \caption{Fit Sample.}
         \label{fig:Fit Sample}
     \end{subfigure}
     \hfill
     \begin{subfigure}{0.45\textwidth}
         \centering
         \includegraphics[width=\textwidth]{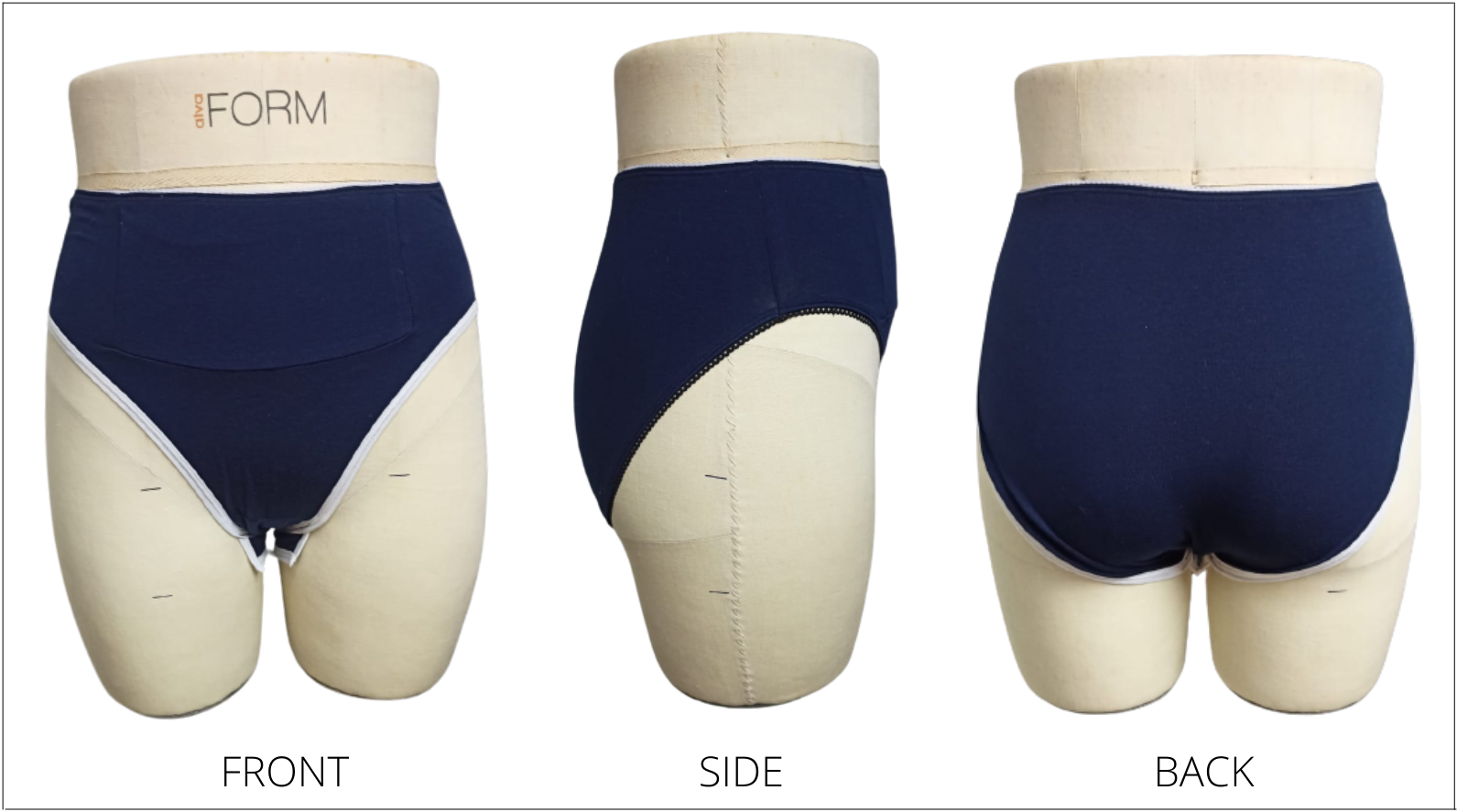}
         \caption{Final Prototype.}
         \label{fig:Final design}
     \end{subfigure}
     \hfill
    \begin{subfigure}{0.45\textwidth}
         \centering
         \includegraphics[width=\textwidth]{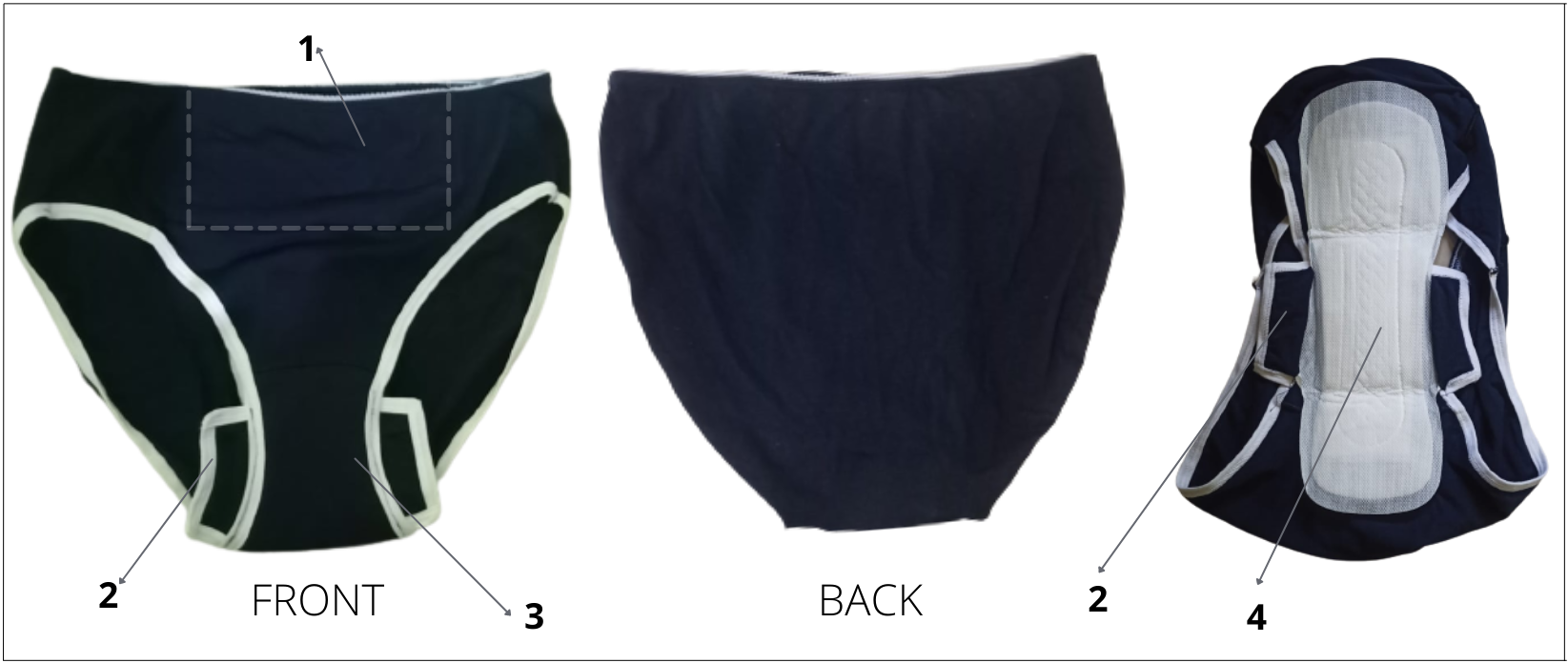}
         \caption{Features included in the garment.}
         \label{fig:garmentexplained}
     \end{subfigure}
     \hfill
     
    \caption{Stages of garment development.}
    \label{stagesgarment}
    \raggedright
\footnotesize{    In Figure \ref{fig:garmentexplained} the components of the garment are shown  and descriptions is mentioned below: 
\begin{enumerate}
    \item The pocket in the abdominal region placed on the front side of MIMA is designed to accommodate the heat pad.
    \item Dedicated wing pockets accommodate the extra pad wings to prevent abrasion between \red{the} skin and the pad\red{-}wing \red{microplastics} thus preventing rashes.
     \item The triple layered leak-proof gusset is designed to prevent any leakage from intimate wear and also stains on external bottom wear.
     \item The sanitary pad must be placed onto the gusset area as referred \red{to} in Figure \ref{fig:garmentexplained}. 
\end{enumerate}
}
\end{figure}

To address leakage issues associated with traditional forms of women's underwear, we incorporated a three-layer gusset panel in MIMA underpants. This panel consists of a leak-proof polyester fleece sandwiched between two layers of garment fabric. Additionally, we identified that using a sanitary pad during menstruation can cause rashes in the inner thigh area, primarily due to the microplastics in the extra pad wings agitating the delicate skin. To mitigate this issue, we designed dedicated pockets in the MIMA pants to accommodate these extra wing pockets, thereby preventing unnecessary contact between the skin and plastic. The garment concept underwent multiple iterations (Figure \ref{stagesgarment}) driven by continuous feedback from our volunteers.

The fabric used in the final garment features an antibacterial coating, which prevents menstrual malodor by inhibiting bacterial growth. Through clinical analysis, we determined that the antibacterial finish remains effective for up to 60 wash cycles. The efficiency of the antibacterial finish gradually declines from 99.4\% to 93.8\% in the first 30 wash cycles, followed by a further decline to 88\% over the next 30 cycles.  \\

\subsection{IoT Integration}
\textcolor{black}{The IoT integration in MIMA aims to enable pain-free periods and reduce the hassles associated with menstruation by incorporating an IoT module into the garment and connecting it to a smartphone app. The app \red{is developed using Flutter SDK \cite{flutter} and} offers additional features such as cycle tracking, reminders, and health tips. We are also working on integrating a 'request a pad' feature, allowing users to request sanitary napkins from nearby app users in emergencies \red{anonymously}. The app provides a platform for future integration of more sensors to provide information about the user's physiology. This comprehensive IoT system aims to provide support, convenience, and personalized insights to enhance the overall menstrual experience. }

IoT system was designed and incorporated as a way to provide an integrated system that can generate safe, controllable, and distributed heating around \red{the} abdominal area of the waistband. We decided on a Bluetooth\red{-}based control system for regulating the heating on \red{the} user's command through a smartphone app  (Figure \ref{fig:screenshot}). The heating system has two components – one is a flexible paper-thin heating pad fabricated out of Nickle Chromium alloy wire and Kepton\textsuperscript{\textregistered} Polyamide tape. This heat pad is connected to a power and control module that is based on a microcontroller and is controlled by the user’s smartphone through Bluetooth. \red{The code for the control module is written in C programming language and compiled onto the Atmega328p microcontroller using Arduino IDE \cite{arduino}.}

\subsubsection{\textcolor{black}{MIMA 1.0 - }The proof of concept}
\begin{figure}[H]
         \centering
         \includegraphics[width=0.6\linewidth]{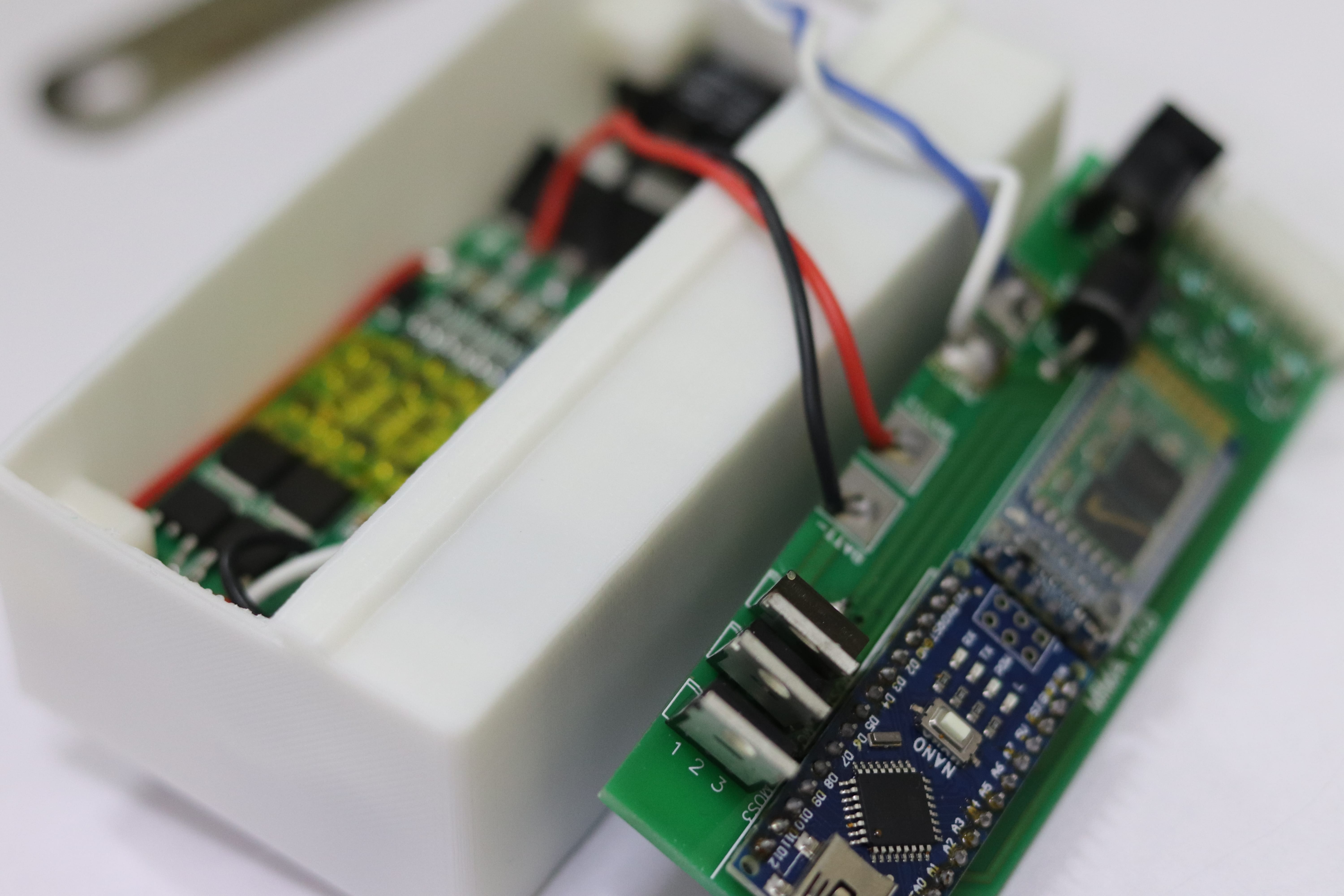}
         \caption{Internal circuitry of the proof of concept.}
         \label{fig:pcb}
     \end{figure}

The system was designed in two stages. In the first stage, we designed a proof of concept (Figure \ref{fig:pcb}) which was based on the control module fabricated using a 3D printed case and off-the-shelf components such as an Arduino Nano, HC-05 Bluetooth module, A 3S 2200 mAh battery\red{,} and an off the shelf BMS board. In this version, the module was designed to be an external unit that was clipped onto the user’s waist over the top layer of clothing. The \red{h}eating pad is a thin sheet of Kepton\textsuperscript{\textregistered} Poly-amide tape with heating elements of Nickle-chromium wire embedded into a 3 coil structure along with dedicated thermistors forming three discrete temperature control zones. This distribution of temperature sensor inputs also forms a way to fail-safe the system against any malfunction in the temperature control system. The heating pad is of dimensions 10 cm x 18 cm and is inserted into a dedicated pocket inside the MIMA period pants.  The control module in the first version was tethered to the \red{h}eating pad via a thin and flexible cable running from the undergarment to the external control module. This \textcolor{black}{proof of concept helped us to gauge the rel\red{e}vance of MIMA as a solution to women during periods. We collected inputs from our volunteers regarding its relev\red{a}nce, and improvements to the design towards a suitable product. Our proof of concept }has since evolved and in\textcolor{black}{to} the current version, \textcolor{black}{wherein} the entire MIMA IoT system is integrated into the undergarment in a removable manner. \textcolor{black}{The specific details and the differences between the proof of concept (MIMA 1.0) and the current version of our product (MIMA 2.0) have been noted in the table \ref{tab:featurecomp} and \ref{tab:feedbackanalysis}. } \\

\begin{figure}[htp]
     \centering
     \includegraphics[width=0.45\linewidth]{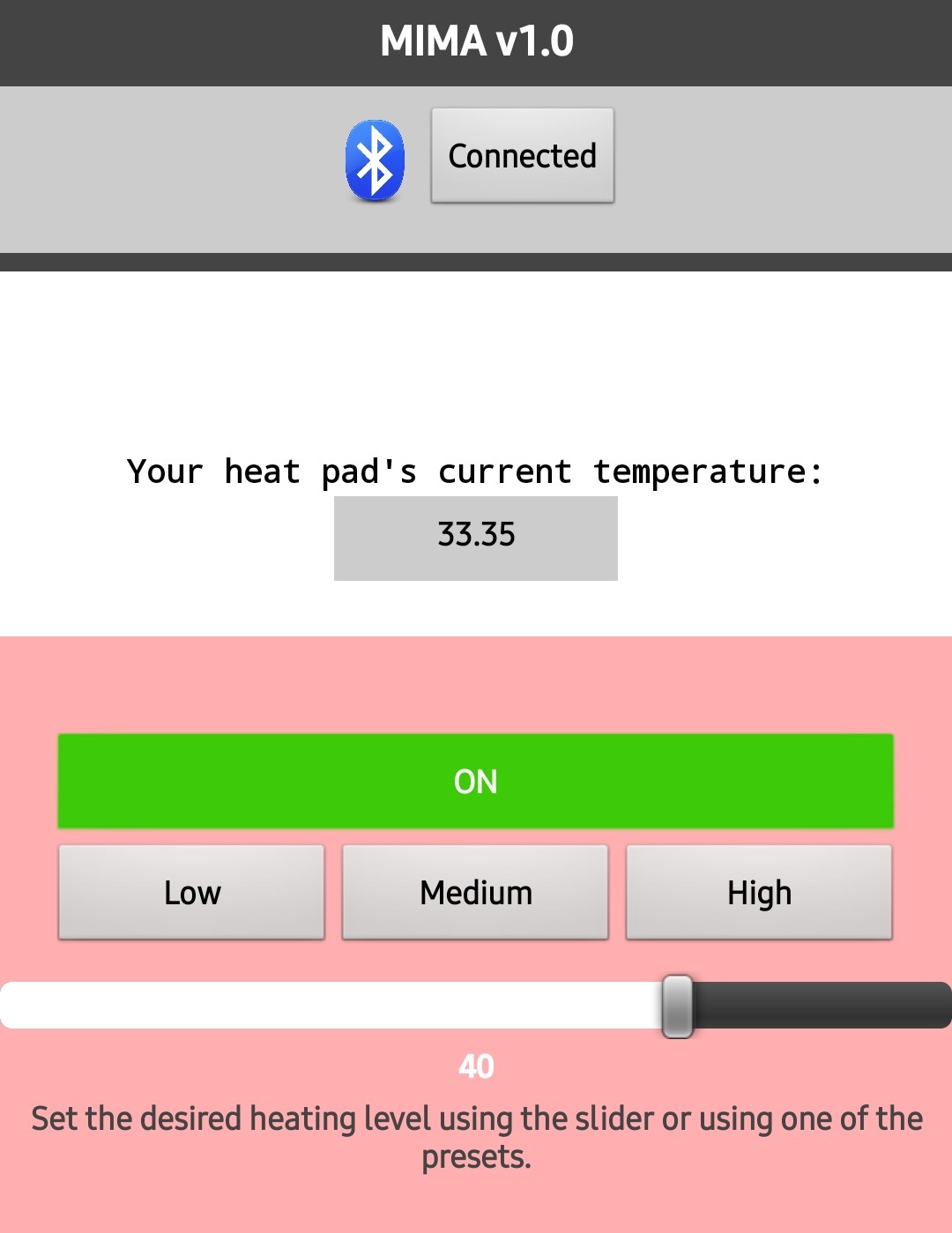}
    \caption{Screenshot of the smartphone app interface in MIMA \cite{mima1}.}
    \label{fig:screenshot}
\end{figure}

The control module connects to the smartphone app and sends temperature sensor values to the app along with receiving control instructions from the app (Figure \ref{fig:screenshot}). The app now also features functions such as manual menstrual cycle tracking, auto timer for heating, etc.

\begin{figure}[H]
     \centering
     
            \centering
            \includegraphics[width=0.8\linewidth]{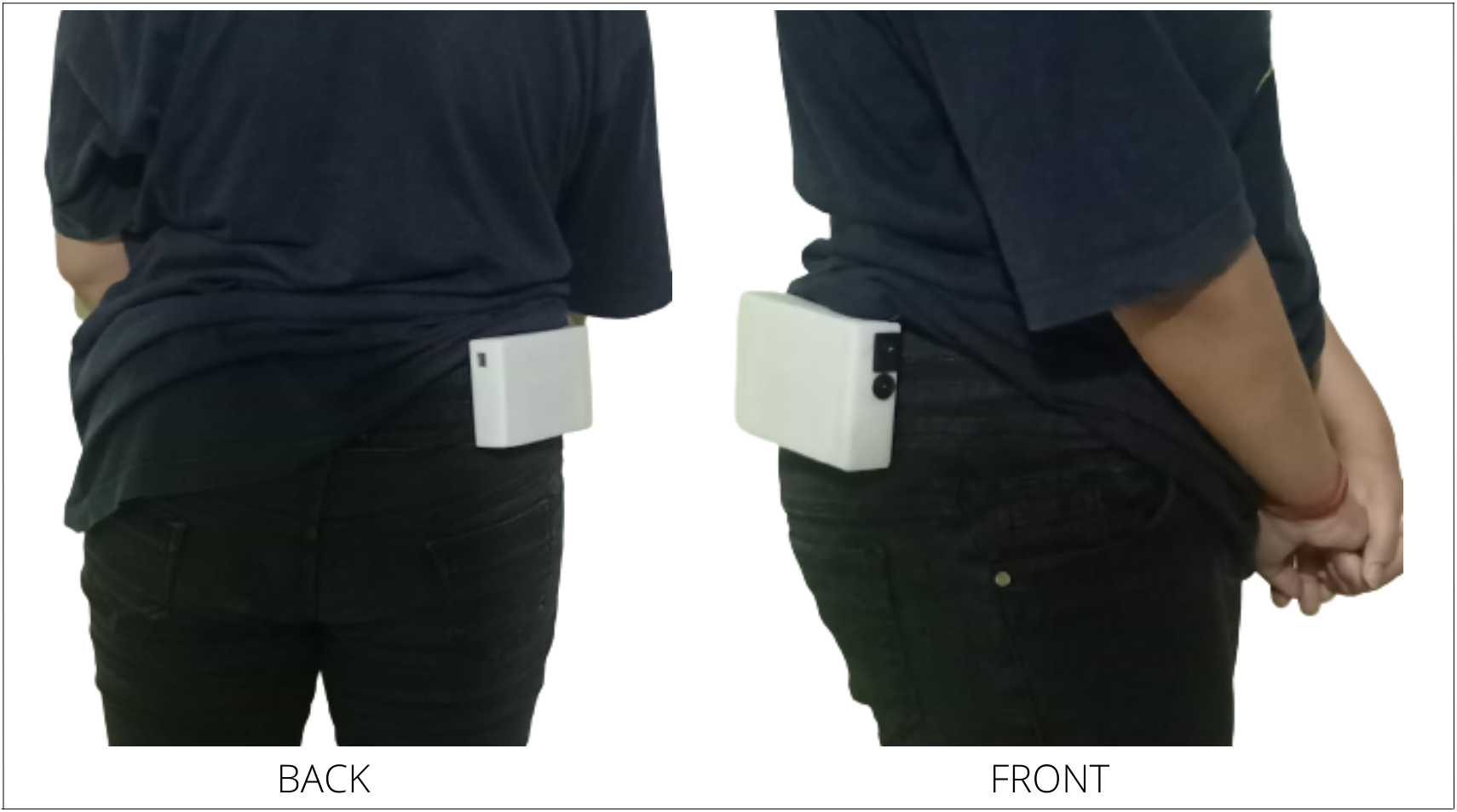}
            \caption{Control module with User.}
            \label{fig:user}
\end{figure}

The proof of concept \textcolor{black}{(MIMA 1.0)} was used to collect feedback from potential users. This helped us understand the need for the product along with a hint of the ideal implementation as desired by the subjects of our survey. The MIMA 1.0 module [\ref{fig:MIMA11-assembly}] used as proof of concept for collecting feedback weighed 178 Grams and had dimensions of 10cm x 6 cm x 3.5cm. The various parts of MIMA 1.0 (Proof of concept) are shown in Figure \ref{fig:MIMA11-assembly} along with \red{the} application of \red{the} external control module in Figure \ref{fig:user}. Figure \textcolor{black}{\ref{fig:mima11assembly}} show\red{s} the dimensions of the MIMA 1.0 control. MIMA's proof of concept was tried by voluntary participants representing all target age groups\red{,} and their feedback was recorded. The volunteers also participated in \red{a} subjective discussion regarding comfort, effectiveness\red{,} and overall experience of the concept.

\begin{figure}[htp!]
     \centering
     \includegraphics[width=0.8\linewidth]{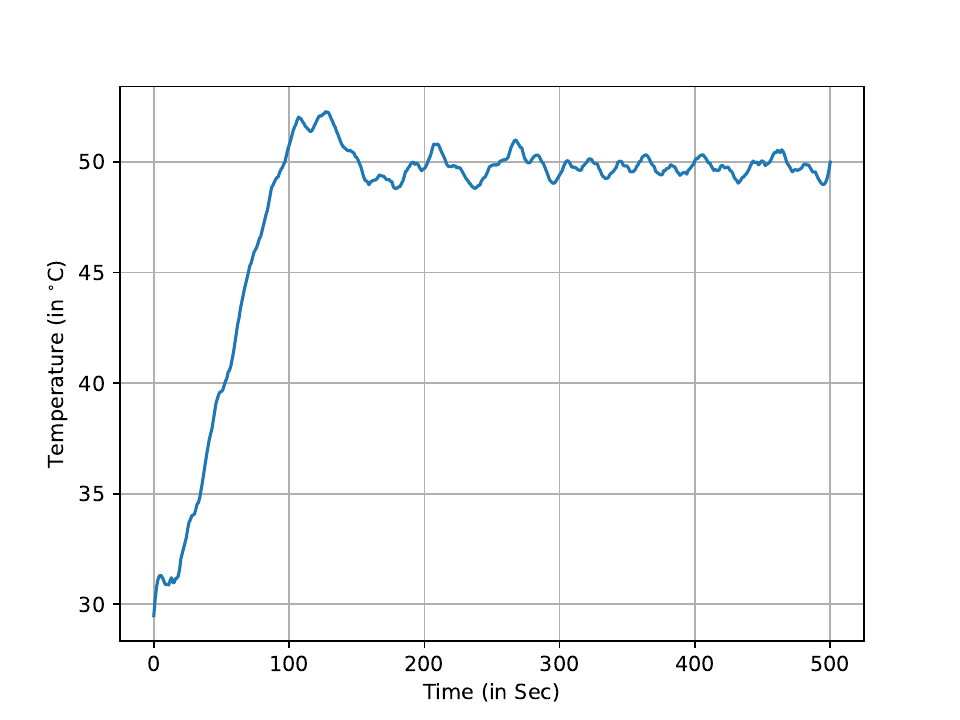}
    \caption{Heating curve of MIMAs IoT Module.}
    \label{fig:heatingcurve}
    
\end{figure}

Under controlled analysis at room temperature (30 degree\red{s} Celsius), The system was able to achieve a peak coil temperature of 55 degree\red{s} Celsius in 90 seconds (Figure \ref{fig:heatingcurve}) while being able to maintain the temperature consistently with a deviation of +/- 0.605 degrees Celsius. It is to be noted that the coil temperature represents the temperature of the actual heating element\red{,} which is shielded within layers of the garment and the Kepton\textsuperscript{\textregistered} Poly-amide material of the heating pad itself. Empirically it is noted that the actual temperature experienced by the user is \~10 degree\red{s} Celsius lower th\red{a}n the coil temperature. \textcolor{black}{ Thus\red{,} the system is designed to provide a comfortable 45-48 degree temperature on the user's skin, which is based on the recommended temperature for heat therapy for menstrual cramps. \cite{ higgins2002quantifying}}

\begin{figure}[H]
     \centering
      \begin{subfigure}{0.45\linewidth}
         \centering
         \includegraphics[width=\linewidth]{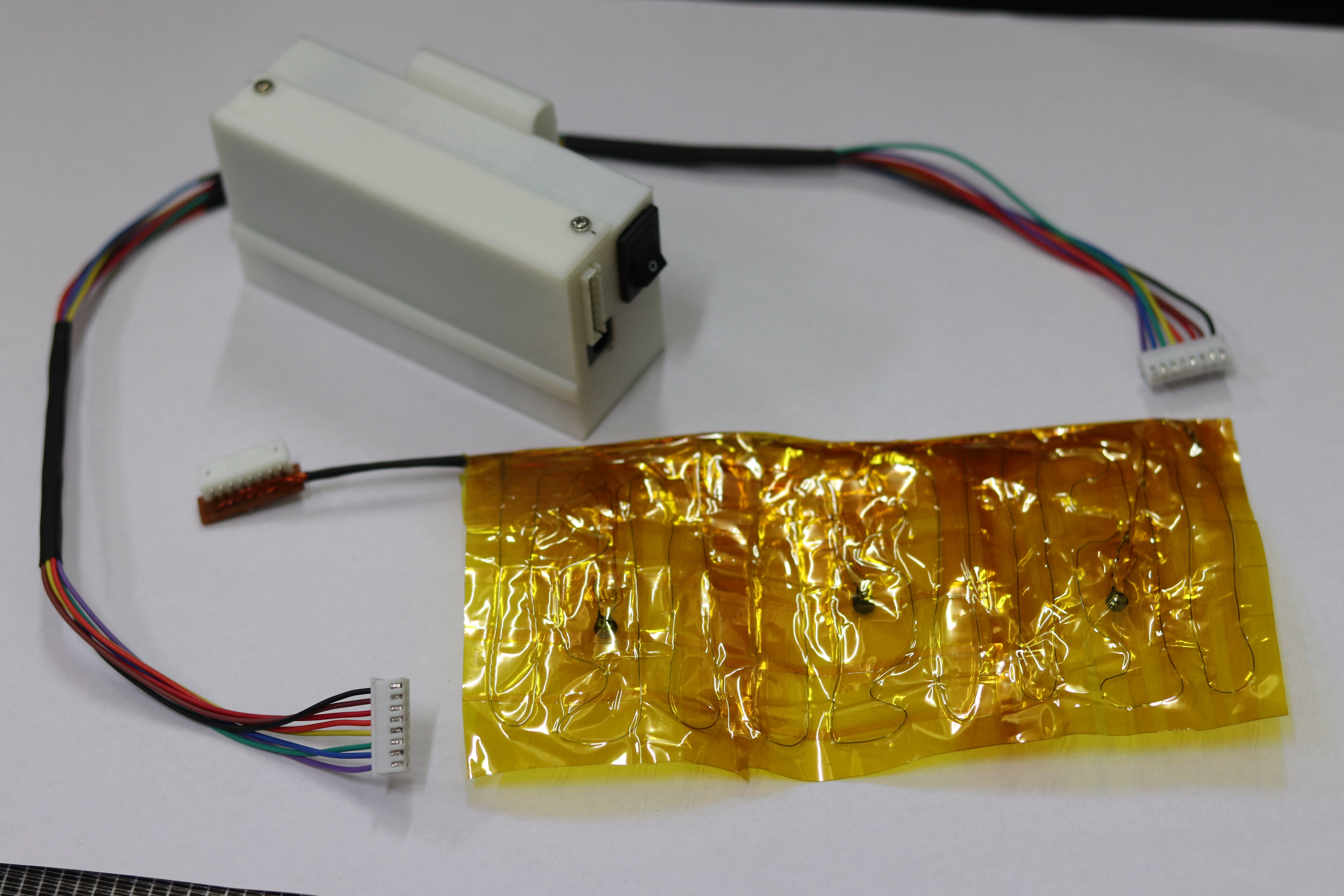}
         \caption{Parts of the system.}
         \label{fig:mima11com}
     \end{subfigure}
     \hfill
       \begin{subfigure}{0.45\linewidth}
         \centering
         \includegraphics[width=\linewidth]{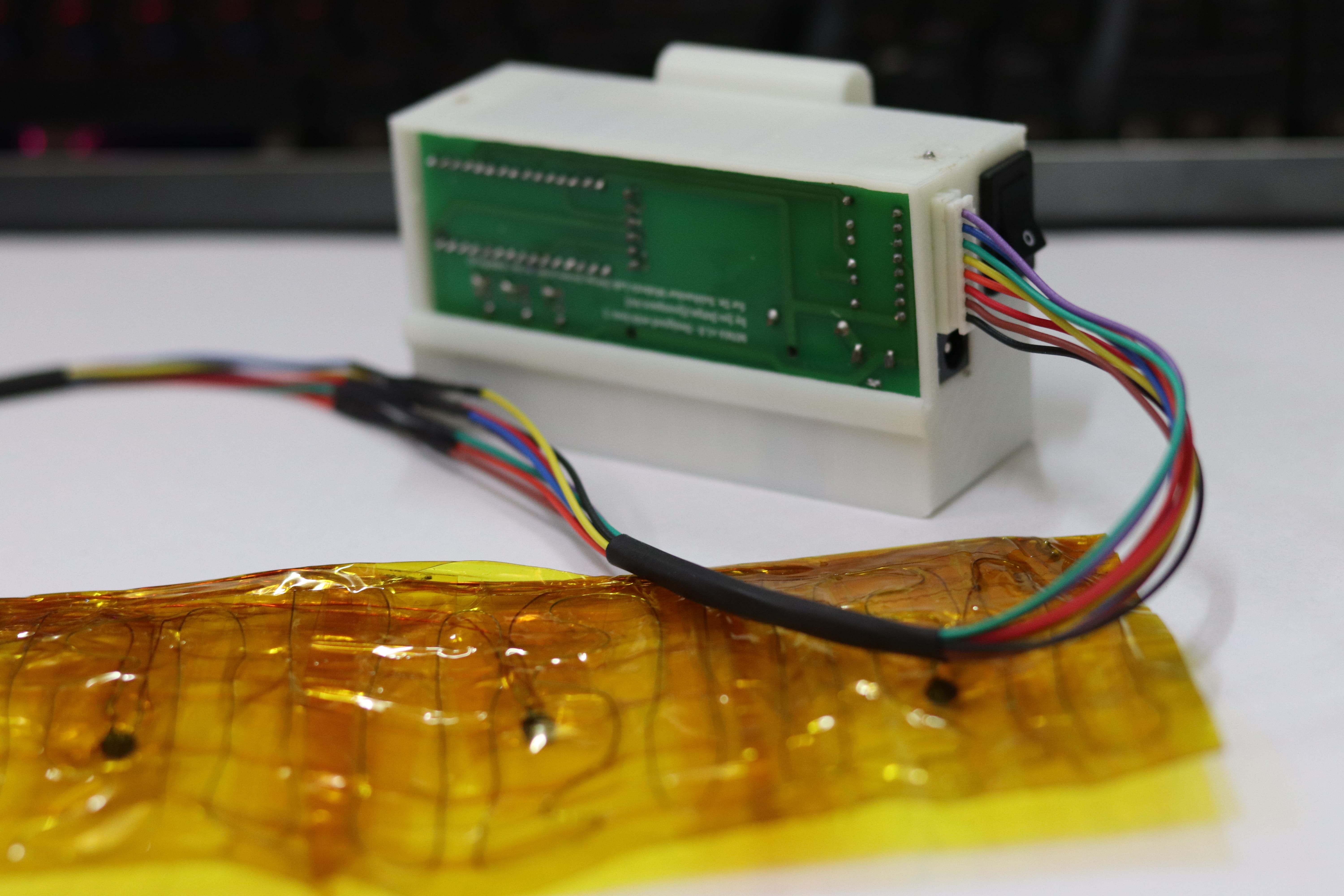}
         \caption{Components assembled.}
         \label{fig:mima11com}
     \end{subfigure}
     \hfill

        \caption{MIMA 1.0 (Proof of concept) IoT System Assembly.}
\label{fig:MIMA11-assembly}
\end{figure}

\begin{figure}
  \ContinuedFloat
\centering
     \begin{subfigure}{0.7\linewidth}
         \centering
         \includegraphics[width=\linewidth]{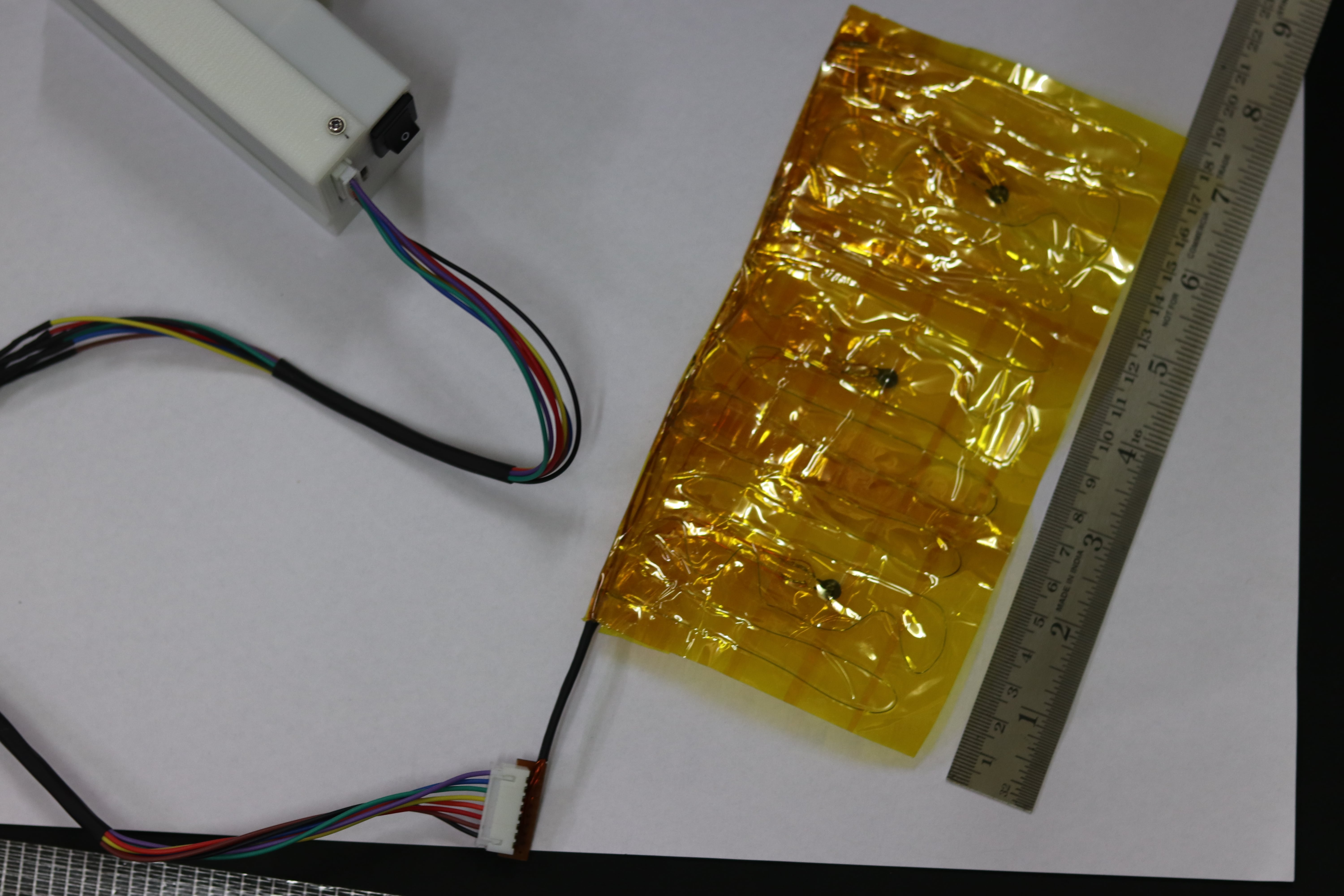}
         \caption{Assembly and a scaled depiction of MIMA 1.0.}
         \label{fig:mima11assembly}
     \end{subfigure}
     \hfill
     
     \caption{MIMA 1.0 (Proof of concept) IoT System Assembly.}
        \label{fig:MIMA11-assembly}
\end{figure}

The firmware of the MIMA control panel is written with various safety features in mind. Multiple checks have been incorporated which ensure a reliable and fail\red{-}safe operation of the heating elements. The code and other resources related to MIMA can be found on our website \cite{mimaweb}. Some of the key features of the system in terms of safe operation are as follows –
\begin{itemize}
    \item All three temperature\red{-}controlled zones should report similar temperature\red{s} to the controller as a significant deviation may indicate a systemic failure. To ensure this, we have programmed the controller to stop heating if the standard deviation of the temperature readings exceeds 2.5. \textcolor{black}{As the standard d\red{e}viation is an absolute value (Given by $s=\sqrt{\sum{(x_{i}-\mu)^2/N}} )$ \red{ where $x_{i}$ is the value of individual sensor, $\mu$ is the mean sensor value, $N$ is the number of sensors (three in our case), and s is the standard deviation of the sensor values. It }reflects the amount of deviation between the values.)}
    \item Maximum coil temperature is capped at 55 degree\red{s} Celsius. In case of an exception to this\red{,} the system cuts off any power to the heating coils. At the maximum coil temperature of 55 degree\red{s} Celsius, the actual\red{ly} felt temperature on \red{the} skin is around 45-48 degree\red{s} Celsius. \textcolor{black}{As suggested in \cite{ higgins2002quantifying}} 
    \item The module must be connected to the smartphone’s Bluetooth at all times to be functional. This prevent\red{s} any mishaps related to \red{the} uncontrolled operation of the heating system.
    \item Bluetooth modules in each of the MIMA IoT control \red{unit is} locked by a secret password to ensure only the owner can connect with the module and thereby control the system.
    \item All anomalies are reported back to the app for \red{the} user's reference.
    \red{, }etc.
\end{itemize} 


Based on the received feedback, we have noted the existing problems in table \ref{tab:feedbackanalysis} along with our proof of concept of MIMA along with our comments towards resolution. 

\begin{table}[htp]
  
\small

\begin{longtable}{|p{0.2\textwidth}|p{0.35\textwidth}|p{0.45\textwidth}|} 
 \caption{Problems analysis from MIMA 1.0 (Proof of Concept).}
 \label{tab:feedbackanalysis}
\\
 
 \hline 
 \textbf{Problem Title}&\textbf{Issue Description}&\textbf{Solution Implemented in MIMA 2.0}\\
 
\hline
  
External Module \ref{fig:MIMA11-assembly} & The external nature of MIMA's IoT module is non-ideal. It would be better if the device was more concealed as menstruation is still regarded as a matter of privacy and secrecy. & MIMA 2.0 module is a tiny unit that gets integrated into the undergarment in a con\red{c}ealed and conv\red{e}nient way.  \\
     
     \hline 
Integration & Some users had difficulty in integrating the heating pad into the clothing. This problem was mainly due to the non-removable tether between \red{the} heat pad and IoT Module in \red{the} initial prototype. &  The MIMA 2.0 module is a composed unit \red{that} can be very easily inserted into the MIMA underpants.   \\

\hline 

Garment Sampling & Sample testing not done for small stature subjects such as adolescent girls who are getting started with their menstrual cycle & We are in \red{the} process \red{of} developing a feasible methodology to produce \red{a} sizable quantity of samples for field\red{-}testing in various size and age group of women.\\

     \hline
     
\end{longtable}
  
  \end{table}

\subsubsection{Feedback analysis \textcolor{black}{leading to the development of MIMA 2.0}}
The responses were collected from our voluntary participants who took trials with our available samples of MIMA 1.0. \red{This} feedback helped us develop our concept and arrive at MIMA 2.0 \ref{subsubsection:mima2}. The feedback collected from the participants who tried our MIMA samples revealed that approximately 60 percent of them found MIMA to be extremely comfortable, appreciating the overall feel of the intimate wear. The design and the choice of a dark shade for the garment were well-received by almost all participants. Around 80 percent of the respondents agreed that the wing pockets added a valuable and comforting functionality to the garment. Additionally, 60 percent of the participants expressed their appreciation for the integrated heat pad feature, with a majority of them preferring a medium range of heat at around 42 degrees Celsius. The app associated with MIMA was reported as user-friendly by all respondents. Sixty percent of the participants found the detachability of the components to be very easy, while the remaining participants were able to attach and detach them easily after trying it more than once. Moreover, 70 percent of the participants liked the ease of clipping the module to the garment. Overall, the experience of using MIMA was described as new, exciting, and highly helpful\red{,} and unique by all participants. The issues reported in this analysis are summarized in table \ref{tab:feedbackanalysis} along with the solution implemented in MIMA 2.0.

\subsubsection{MIMA 2.0}
\label{subsubsection:mima2}

\begin{figure}[h!]

     \centering
      \begin{subfigure}[H]{\linewidth}
         \centering
         \includegraphics[width=0.8\linewidth]{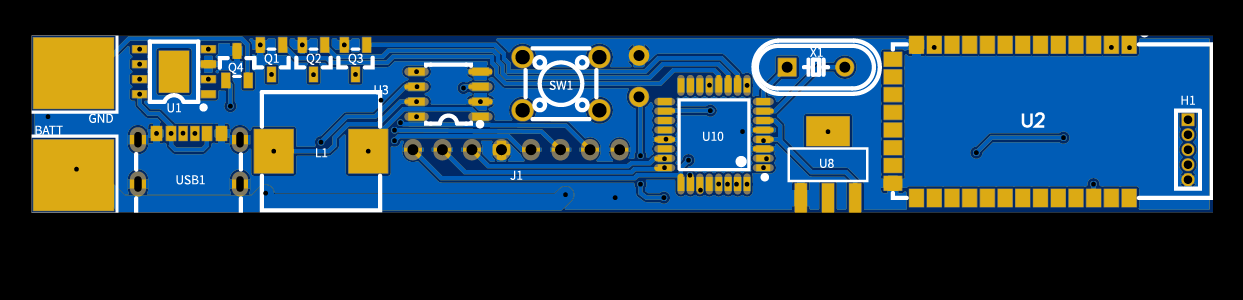}
         \caption{Front.}
         \label{fig:pcbschmima2front}
     \end{subfigure}
    \begin{subfigure}[H]{\linewidth}
         \centering
         \includegraphics[width=0.8\linewidth]{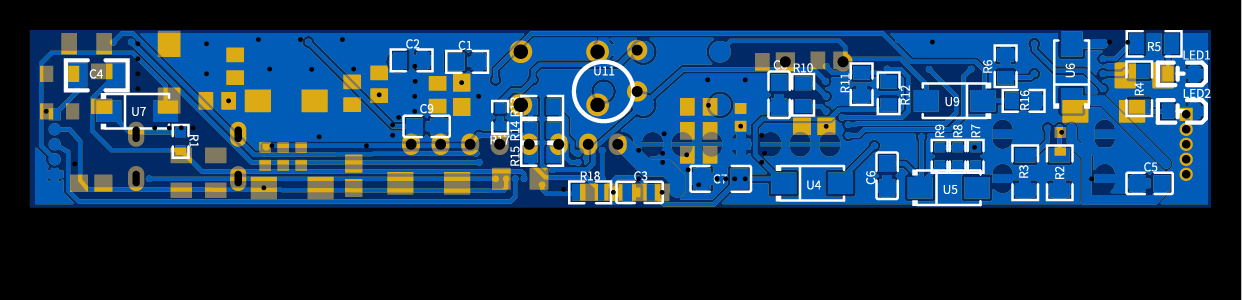}
         \caption{Back.}
         \label{fig:pcbschmima2front}
     \end{subfigure}
     \caption{MIMA 2.0 PCB Schematic.}
    \label{fig:PCB_New}
\end{figure}

\begin{figure}[h!]
     \centering
     \includegraphics[width=\linewidth]{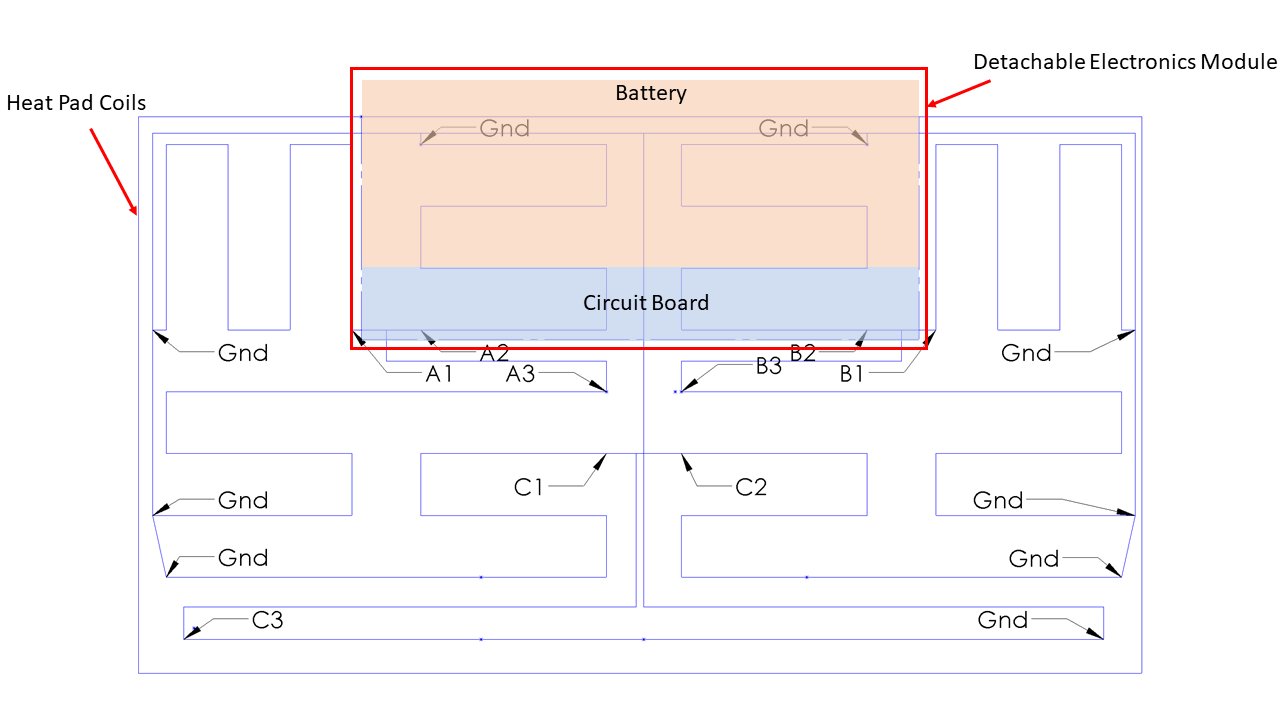}
    \caption{Layout of MIMA IoT Unit.}
    \label{fig:MIMA2Layout}
    
\end{figure}

\begin{figure}[h!]
     \centering
     \includegraphics[width=\linewidth]{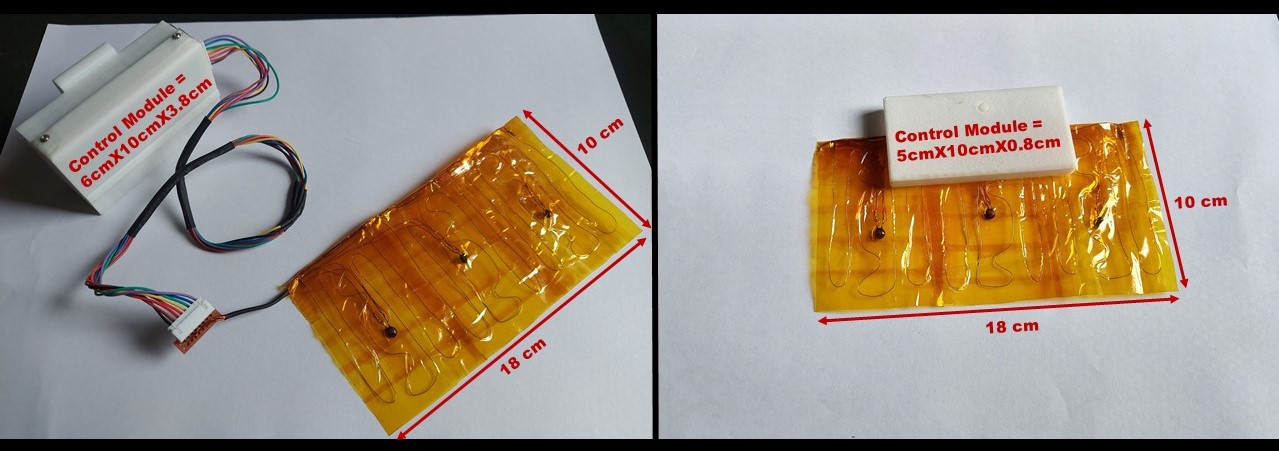}
    \caption{Comparison between Version 1 and Version 2.}
    \label{fig:mimacomp}
    
\end{figure}

Addressing the issues pointed out in the feedback towards our proof of concept, We worked towards further development with a focus on miniaturizing and integrating the system within the undergarment in a removable, safe, and concealed manner (Figure \ref{fig:MIMA2Placement}). A detailed feature comparison of MIMA 1.0 and MIMA 2.0 is given in table \ref{tab:featurecomp}, and a visual comparison is given in Figure \ref{fig:mimacomp}. We used an ATMega328p microcontroller chip for controlling the system and developed a fresh 3.7-volt system at a component level (Figure \ref{fig:PCB_New}). The new system runs on a single 2200mah LiPo cell which is fitted with the module temperature and other safety parameters in mind. This helped us in making the system very thin (7.5 mm) and small in size (5cmx 10cm). The heating pad is composed of a Kepton polyamide base and 9 heating coils made out of 36 AWG \red{N}ichrome wire (Figure \ref{fig:MIMA2Layout}). The heating coils are distributed into three temperature zones.  The heating pad and the control unit are attached with a detachable 8-pin JST-XH connector. The control module in turn is attached to the inside of the garment using \red{V}elcro on the garment layer away from the user’s skin. 

We would like to accept that given the current state of technology, with the current version, we are at the limit of the miniaturization of this product. The bulk of the product is constituted by the LiPo battery and reducing size further would require reducing the battery capacity. The current can run on full power (~48 degrees Celcius) constantly for around 30 minutes depending on insulation provided by external clothing, ambient temperature etc. However, in a practical use case, a medium setting (~42 degrees Celcius) is recommended in stretches of 5-8 minutes of heat application per session. This ends up extending the battery life to practically 6-8 hours which reflects a typical work day for the majority of our target audience. The product holds the potential to particularly help working women and young school-going girls with painful menstrual cramps.  We subjected the current version through a similar survey and feedback analysis as we did with our prototype and received a very positive response from all of our reviewers. The current version may be deemed a viable product as a menstrual aid for women. A comparative analysis for MIMA and MIMA 2.0 is given in table \ref{tab:MIMAComp} The feedback analysis for MIMA 2.0 is given in table \ref{tab:FeedbackMIMA2}. 














\begin{table}[H]
\caption{Comparative analysis of MIMA 1.0 and MIMA 2.0 features.}
\label{tab:MIMAComp}
\begin{tabular}{|l|ll|}
\hline
\textbf{Feedback Criteria} & \multicolumn{2}{l|}{\textbf{\begin{tabular}[c]{@{}l@{}}Comparison of responses between\\ 1st \& 2nd Prototype\end{tabular}}} \\ \hline
Demography & \multicolumn{2}{l|}{\begin{tabular}[c]{@{}l@{}}MIMA was tried by age 17-42 years \\ \& have helped with the responses\end{tabular}} \\ \hline
Comfort of MIMA & \multicolumn{2}{l|}{\begin{tabular}[c]{@{}l@{}}The 2nd Prototype was more \\ comfortable because of comparatively \\ more flexible connections and slightly \\ more compact module\end{tabular}} \\ \hline
Ease of Clipping & \multicolumn{2}{l|}{\begin{tabular}[c]{@{}l@{}}The method of clipping remains the \\ same but the 2nd prototype feels \\ comparatively comfortable due to \\ slight difference in shape \& weight \\ of the control module\end{tabular}} \\ \hline
Overall feel of Intimate Wear & \multicolumn{2}{l|}{\begin{tabular}[c]{@{}l@{}}Most of the women feel very \\ comfortable in MIMA\end{tabular}} \\ \hline
\begin{tabular}[c]{@{}l@{}}Weight \& Compactness of the\\  control module\end{tabular} & \multicolumn{2}{l|}{\begin{tabular}[c]{@{}l@{}}There might be slight difference \\ in the weight but the overall module \\ is found to be more compact in terms \\ of shape and detachability of the \\ connections which makes it more \\ handy than before\end{tabular}} \\ \hline

\end{tabular}

\end{table}

\section{Result and Analysis}
\label{sec:result}
MIMA represents a pioneering approach to providing women with a functional and adaptable menstrual aid that significantly improves comfort and convenience. It specifically caters to the needs of school-going adolescent girls and working women who often face challenges due to painful cramps and other side effects of menstruation. By wearing MIMA, these individuals can now experience enhanced comfort while continuing with their studies or work without any compromise. MIMA offers a smart, integrated, concealed, and comfortable intimate wear solution that aligns perfectly with the demands of modern lifestyles. With its advanced features, MIMA is poised to revolutionize the experience of menstruation for women, empowering them to navigate their daily lives with greater ease and confidence. \\

The MIMA IoT module is seamlessly integrated into an internal pocket within the MIMA period pants, ensuring a discreet and convenient experience for the user. To ensure the product's efficacy and user satisfaction, extensive trials and feedback sessions were conducted with participants who actively tested and provided their valuable input on the comfort and overall experience of using MIMA. The feedback received from the participants has been thoroughly analyzed and summarized in the previous section. This iterative process allowed us to refine and enhance the product at each stage of development, ensuring that it meets the needs and expectations of our target audience. The feedback analysis was conducted with the same group of participants who have been involved and consulted throughout the entire development process, starting from the initial proof of concept. Their valuable insights and opinions have been instrumental in shaping the final version of MIMA, making it a truly user-centric solution to menstrual problems. As concluded \red{in} table \ref{tab:FeedbackMIMA2}, it was observed that there was a noticeable improvement in comfort and utility of the control module due to its concealed nature, and a significant reduction in weight along with much easier detach-ability. \\

As noted earlier, the heating system \red{can} achieve a comfortable level of warmth very quickly on demand. As depicted in the heating curve (Figure \textcolor{black}{\ref{fig:heatingcurve}}) the system was able to achieve a maximum coil temperature of 55 degree\red{s} Celsius (\textcolor{black}{\~ 45-48} degree Celsius felt on skin) within 90 seconds and was able to maintain a consistent temperature within (2.5) degree Celsius.  The actual temperature felt on \red{the} skin is around 10 degree\red{s} Celsius lower than the coil temperature due to insulating layers of clothing and Kepton\textsuperscript{\textregistered} Polyamide base of the heating pad. \textcolor{black}{The system consistently checks the temperature measurements two times a second.} \\

The fabric of MIMA period pants features an antibacterial coating \red{that} prevents \red{the} growth of bacteria \emph{Gardenerella vaginalis} which is primarily responsible for the unpleasant fishy smell often referred to as menstrual malodor. The antibacterial coating was clinically tested to have an efficiency of 99.4 percent in the first use with \red{a} gradual decline to 93.8 percent after the first 30 wash cycles and then to 88 percent at the 60\red{-}wash cycle mark. The antibacterial coating was seen to be effective in practice in eliminating menstrual odor considerably.\\

The approach taken by MIMA to redesign traditional underwear specifically for menstruation has received positive feedback from the majority of survey participants. The final analysis of the feedback validates our approach, as many features such as the pad wing pockets, leak-proof gusset, and the inclusion of a heating device while maintaining privacy were highly appreciated by the users. These features address common concerns and challenges faced by women during menstruation, providing them with practical solutions and enhancing their overall experience. The positive response from the users reinforces the effectiveness and value of MIMA's innovative design and functionality.\\

\textcolor{black}{As concluded in the table \ref{tab:MIMA Cost}, MIMA was carefully designed with a vision of hassle-free menstruation for women \red{a}round the globe with unique features, re-usability\red{,} and affordable price-point,  making it a great value for money. As discussed in the table \textcolor{black}{\ref{tab:MIMA_durability}}, Durability and Sustainability were the key aspects considered \red{during} the product development.  \\ }

\begin{table}[]
    \centering
    \begin{tabular}{|c|c|c|}
    
    \hline
         \textcolor{black}{\textbf{Expense} }& \textcolor{black}{ \textbf{Amount}} & \textcolor{black}{ \textbf{Type}} \\
         \hline
        \textcolor{black}{Control Module} & \textcolor{black}{2,000 INR (25.3 USD)} & \textcolor{black}{One time purchase} \\
        \hline
        \textcolor{black}{Individual garment} & \textcolor{black}{650 INR (8.2 USD)} & \textcolor{black}{Recurrent Expense} \\
        \hline
        \textcolor{black}{Total Cost} & \textcolor{black}{2,650 INR (33.1 USD)} & \textcolor{black}{First time purchase cost} \\
        \hline 
    \end{tabular}
    \caption{\textcolor{black}{Cost analysis of MIMA 2.0}.} 
    \label{tab:MIMA Cost}
\end{table}

\begin{table}[]
    \centering
    \begin{tabular}{|c|c|}
    \hline
        \textcolor{black}{\textbf{Number of Cycles}} & \textcolor{black}{\textbf{Antibacterial finish effectiveness}} \\
        \hline
       \textcolor{black}{ Upto 30 cycles} & \textcolor{black}{99.4 \% to 93.8 \%} \\
        \hline
        \textcolor{black}{30 to 60 cycles} & \textcolor{black}{95 \% to 88 \%} \\
        \hline
        
    \end{tabular}
    \caption{\textcolor{black}{Wash durability of MIMA 2.0: Antibacterial finish works effectively to up to 60 laundry cycles.}}
    \label{tab:MIMA_durability}
\end{table}

\begin{table}[htp]
    \centering
   \caption{Feedback analysis for MIMA 2.0.}
   \label{tab:FeedbackMIMA2}

\begin{longtable}{|p{0.2\textwidth}|p{0.35\textwidth}|p{0.35\textwidth}|}

    \hline
    \textbf{Feature}& \textbf{Response}&\textbf{Remark}\\
     \hline
     \endhead
Prior familiarity with MIMA & Most of our subjects have used MIMA 1.0 in past. Most of them had received the concept well and some provided critical feedback that helped us provide improvements in MIMA 2.0 & Prior familiarity of the subjects allowed for reviews that were more directed towards changes/ improvements to the concept. Majority of the critical reviewers from past reported satisfaction with the current version \\

\hline
Placement of the Heat-pad and control module & Positioning of heat-pad as well as the control module was reported ideal by almost all respondents. & The incorporation of the control module into the garment itself allows for a private and concealed usage which was received well by majority of the subjects \\

\hline
Utility of the control module & All respondents consider the new module lighter, compact and easier to incorporate compared to MIMA 2.0. & Acceptance of MIMA as a viable product that they'd like to use as their usual menstrual aid has increased. \\

\hline
Ease of ataching and detaching the module & 80 \% of the respondents could easily attach and detach the heat pad and the control module in the pocket. & Others believe that attaching and detaching was easier and faster after few repetitions. \\

\hline
Overall experience of MIMA 2.0 & 60\% of the respondents feel comfortable in adopting the current version of MIMA as their menstrual aid. & Others feel it is comfortable but it is a new experience \& agree that they could get comfortable with it with prolonged usage. \\
\hline

\end{longtable}

\end{table}

\section{Conclusion and Future Work}
\label{sec:conclusion} 
\subsection{Conclusion}
\red{MIMA 2.0 is the result of ongoing efforts based on feedback and lessons learned from various trials involving voluntary participants throughout our product development journey. We conducted both quantitative and qualitative analyses after each phase, consistently upgrading the module. This iterative process has led to the creation of an enhanced heating and circuit system. The improved version now features a sleeker heat pad integration, paired with a remarkably compact and lightweight control module. This design upgrade ensures easier detachability and provides a more comfortable experience for our users. The positive response from participants has been overwhelming, indicating their appreciation for the improvements made.}

\subsection{Future Work}
\red{Our future efforts will primarily focus on enhancing the cost-effectiveness of MIMA, while also striving to create a thoughtfully designed version with additional features to address the specific needs of girls experiencing early menarche, particularly in the age group of 9 years and older. Presently available products do not cater specifically to their requirements. Our goal is to develop a comprehensive product that addresses multiple issues women face during their menstrual cycle.} \\

\red{We are also exploring improvements to the existing version of MIMA, such as incorporating a similar heating system for the user's back, exploring better and lighter alternatives for the existing battery technology, and refining the design and integration into the garment. Additionally, we are considering the incorporation of non-contact physiological sensors, such as pulse rate monitors, pulse oximeters, and IR-based temperature monitoring, into the control module. This innovation could enable the close monitoring of vitals around the user's intimate area, potentially leading to AI-based estimations of various parameters, including stress levels, hormonal status, infections, abnormalities in the cycle, and predictions of cramps, among other factors.}

\bibliography{mybibfile}

\end{document}